\documentclass[suppldata]{interact}

\usepackage{epstopdf}
\usepackage{subfigure}

\usepackage[english]{babel}
\usepackage[
    natbib=true,
    style=numeric,
    sorting=none
]{biblatex}

\usepackage{amsmath,amssymb,amsfonts}
\usepackage{mathtools}
\usepackage{algorithmic}
\usepackage{graphicx}
\usepackage{textcomp}
\usepackage[bookmarks=false]{hyperref} 
\usepackage[export]{adjustbox}
\usepackage{booktabs}
\usepackage{multirow}
\usepackage{multicol}
\usepackage[flushleft]{threeparttable}
\usepackage{soul}
\usepackage{enumitem}
\usepackage{frcursive}
\usepackage{csquotes}

\usepackage{url}
\usepackage[version=3]{mhchem}
\usepackage{longtable}

\usepackage{footnote}
\makesavenoteenv{tabular}
\makesavenoteenv{table}
\usepackage{enumitem}
\usepackage{ragged2e}
\usepackage{enumitem}

\newlist{questions}{enumerate}{2}
\setlist[questions,1]{label=\textbf{RQ}\textbf{\arabic*}\textbf{.},ref=\textbf{RQ}\arabic*}
\setlist[questions,2]{label=\textbf{(\alph*)},ref=\textbf{\thequestionsi(\alph*)}}
\usepackage{tabularx}
    \newcolumntype{L}{>{\raggedright\arraybackslash}X}
\begin{document}
\articletype{Journal Article}

\title{A Systematic Review of Online Exams Solutions in E-learning: Techniques, Tools and Global Adoption}

\author{
\name{Abdul Wahab Muzaffar\textsuperscript{1}, Muhammad Tahir\textsuperscript{1}, Muhammad Waseem Anwar\textsuperscript{2,*}, Qaiser Chaudry\textsuperscript{2}, Shamaila Rasheed Mir\textsuperscript{3} and Yawar Rasheed\textsuperscript{2}}
\affil{\textsuperscript{1} College of Computing and Informatics, Saudi Electronic University, Riyadh, Saudi Arabia} \affil{\textsuperscript{2} College of Electrical and Mechanical Engineering, National University of Sciences and Technology (NUST), Islamabad, Pakistan}
\affil{\textsuperscript{3}College of Computer Science and Engineering, University of Hail, Saudi Arabia}
\vspace{0.4cm}
Correspondence should be addressed to Muhammad Waseem Anwar; waseemanwar@ceme.nust.edu.pk
}

\maketitle

\begin{abstract}
E-learning in higher education is exponentially increased during the past decade due to its inevitable benefits in critical situations like natural disasters (e.g. COVID-19 pandemic etc.) and war circumstances. The reliable, fair, and seamless execution of online exams in E-learning is highly significant. Particularly, online exams are conducted on E-learning platforms without the physical presence of students and instructors at the same place. This poses several issues like integrity and security during online exams. To address such issues, researchers frequently proposed different techniques and tools. However, a study summarizing and analyzing latest developments, particularly in the area of online examination, is hard to find in the literature. In this article, a Systematic Literature Review (SLR) of online examination is performed to select and analyze 53 studies published during the last five years (i.e. Jan 2016 to July 2020). Subsequently, five leading online exams features targeted in the selected studies are identified. Moreover, underlying development approaches for the implementation of online exams solutions are explored. Furthermore, 16 important techniques / algorithms and 11 datasets are presented. In addition to this, 21 online exams tools proposed in the selected studies are identified. Additionally, 25 leading existing tools used in the selected studies are also presented. Finally, the participation of countries in online exam research is investigated. Key factors for the global adoption of online exams are identified and compared with major online exams features. This facilitates the selection of right online exam system for a particular country on the basis of existing E-learning infrastructure and overall cost. To conclude, the findings of this article provide a solid platform for the researchers and practitioners of the domain to select appropriate features along with underlying development approaches, tools, and techniques for the implementation of a particular online exams solution as per given requirements.
\end{abstract}


\section{Introduction}\label{intro}
E-learning has shown promising results during critical circumstances like natural disasters, wars, and pandemics like COVID 2019. For that reason, numerous methodologies and learning management systems have been introduced during the last three decades in order to deliver and promote E-learning successfully \cite{mccoy2015author}. The usage of E-learning is continuously growing as yet, which creates opportunities as well as challenges from online lecture delivery, content management, and handling the online exams effectively. Particularly, different technological advancements with reliable and high-speed internet infrastructure allows the exploitation of advanced image processing and machine learning techniques for the realistic accomplishment of educational activities through E-learning~\cite{abisado2019modeling}. This urges colleges and universities for the adoption of E-learning as a reliable educational platform. 

Online examination is an integral part of E-learning solutions for the genuine and fair assessment of students’ performance~\cite{ andersen2020adapting}. The design and execution of online exams are the most challenging aspects in E-learning. Particularly, online exams are usually conducted on E-learning platforms without the physical presence of students and instructors at the same place. This creates several loopholes in terms of integrity and security, of online exams. For example, the verification of an examinee is extremely problematic in online environment particularly in the absence of continuous monitoring. Moreover, online exam settings are highly supportive for cheating as thousands of online information resources are accessible to students without any check and balance. Furthermore, it is very difficult to ensure the high speed and continuous availability of internet connection for all students during exams. The development of effective question banks, impartial setting of exam papers and marking of descriptive questions are few more challenges in online exams. All aforementioned issues eventually compromise the integrity, security, and objectivity of online exams.   

To confront the concerns accompanied by online exams, researchers frequently propose different solutions. Particularly, the online exams features like examinee verification\mbox{~\cite{4-karim2015review}}, abnormal behavior detection, security of overall system, question bank generation etc. are highly important. To improve such online exams features, researchers utilized different development techniques like machine learning / artificial intelligence\mbox{~\cite{5-alrubaish2019automated}}, formal methods etc. Moreover, different datasets have been developed for the evaluation of online exams techniques. Furthermore, several tools have been developed (e.g. Secure Exam Environment\mbox{~\cite{7-frankl2011secure}} etc.) for the efficient online exam execution. The proposed techniques and tools certainly improved the integrity, security and fairness of online exams.
In literature, there exist several studies~\cite{mccoy2015author,chang2016review} where intensive reviews and surveys are performed for the investigation of E-learning as a whole. On the other hand, there are few attempts to analyze particular aspects of online examination like user authentication~\cite{4-karim2015review}, relationship to student learning~\cite{boitshwarelo2017envisioning} etc. Furthermore, a latest systematic review is also available~\cite{6-Kerryn2020} where significant online exams themes like student performances, perceptions, anxiety level etc. are thoroughly investigated. However, a study systematically analyzing and summarizing across-the-board online exams developments is hard to find in the literature to the best of our knowledge. As online exam is a critical part of E-learning, it is a need of the day to investigate and summarize the latest online exams progress like important features, underlying development techniques, tools, datasets and global adoption factors is hard to find in the literature to the best of our knowledge. As online exam is a critical part of E-learning, it is a need of the day to investigate and summarize the latest online exams progress within a single study. To achieve this, a Systematic Literature Review (SLR)~\cite{kitchenham2004procedures} is performed in this article to find the answers of the following questions:
\begin{questions}
        \item \emph{What are the leading studies particularly dealing with the online exams solutions during the past five years i.e. 2016-to-2020?}
        \item \emph{What are the major online exams features reported in the literature during the past five years?}
        \item \emph{What are the main underlying development approaches that have been employed for the implementation of online exams solutions?}
        \item \emph{What are the leading techniques / algorithms proposed in the domain of online exams?}
        \item \emph{What are the major online exams tools proposed in the literature and how existing tools are utilized in the online exams research?}
        \item \emph{What are the leading datasets proposed / utilized for the online exams solutions?}
        \item \emph{What are the main countries contributed / participated in the online exam research?}
        \item \emph{What are the key factors towards the global adoption of online exams and how to promote it in different countries with varying E-learning infrastructure and financial requirements?}
        \item \emph{What are the major challenges in current online exams research and how to improve upon these challenges?}
        \label{itm:qwithlabel}
\end{questions}

\begin{figure}[hbt!]
    \centering
    \includegraphics[width=\textwidth,]{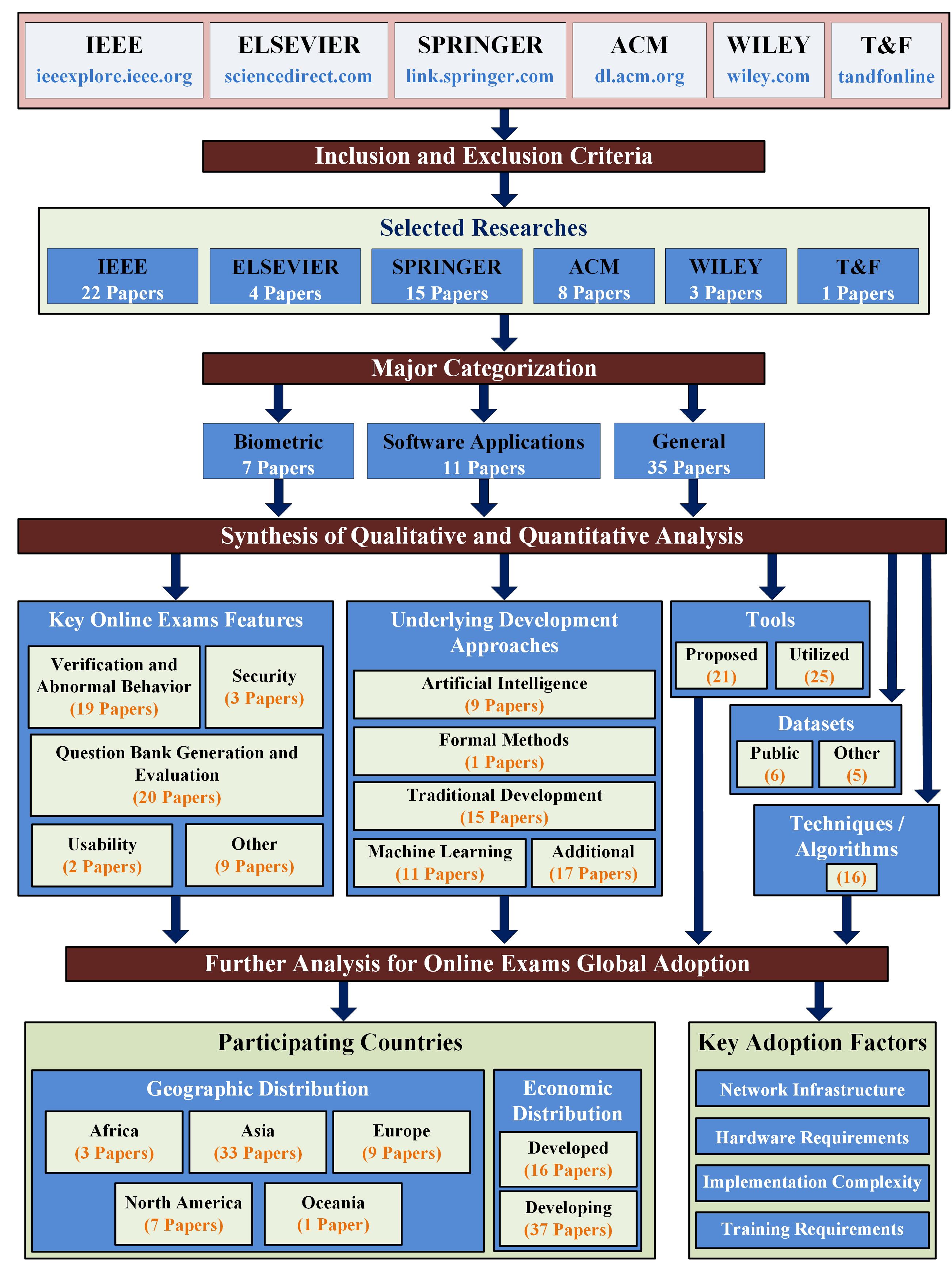}
    \caption{Outline of Research.}
    \label{fig1:Research_Outline}
\end{figure}

To answer aforementioned questions, this article performs SLR to select and analyze 53 studies~\cite{asep2019design,aisyah2018development,natawiguna2016virtualization,ghizlane2019new,ketui2016item,atoum2017automated,sukmandhani2019face,cote2016video,sukadarmika2018introducing,prathish2016intelligent,hu2018research,garg2020convolutional,vomvyras2019exam,matveev2020virtual,chua2019online,mathapati2017secure,shi2017research,mahatme2016data,fan2016gesture,das2019examination,boussakuk2019online,lemantara2018prototype,abisado2018towards,chen2018application,wagstaff2019automatic,rajala2016automatically,opgen2018application,zhang2018analysis,sultan2019automatically,fanani2019interactive,traore2017ensuring,subramanian2018using,frankl2018pathways,sabbah2017security,ettarres2017evaluation,tashu2019intelligent,kolhar2018online,kassem2017formal,jiang2019design,nandini2020automatic,albastroiu2018exam,kausar2020fog,kar2017novel,ullah2019dynamic,diedenhofen2017pagefocus,golden2020addressing,karim2016proposed,wu2020exam,manoharan2019cheat,d2017conceptual,al2019integrated,baykasouglu2018process,chuang2017detecting}, which are published during Jan 2016 to July 2020. The outline of SLR is shown in Fig. \ref{fig1:Research_Outline}. Particularly, a review protocol is developed (Section \ref{reviewProtocol}) with inclusion and exclusion criteria (Section \ref{InclusionExclusionCriteria}) for the execution of this SLR. Six renowned databases (i.e. IEEE, Elsevier, Springer, ACM, Wiley and Taylor \& Francis) are considered for the selection of relevant studies on the basis of inclusion and exclusion criteria. Consequently, 53 studies are selected and classified into three major categories (Section \ref{category}) i.e. Biometric (7 studies), Software Applications (11 studies) and General (35 studies). Subsequently, the combination of both qualitative and quantitative analysis is performed on selected studies to obtain the required and précised results (Section \ref{result}).

It is evident from Fig. \ref{fig1:Research_Outline} that we identify five leading online exam features (Section \ref{leadingAttribute}) that have been targeted in the selected studies. Moreover, we also identified underlying development approaches (Section \ref{developmentApproach}) for online exams through five major class’s i.e. Machine Learning (11 studies), Artificial Intelligence (9 studies), Formal Methods (1 study), Traditional Development (15 studies) and Additional (17 studies). Furthermore, 21 tools proposed in the selected studies with different features are identified and analyzed in Section \ref{proposedtool}. Additionally, 25 significant existing tools utilized in the selected studies for different purposes are also recognized in Section \ref{utilizedtool}. In addition to the tools, 16 noteworthy techniques / algorithms and 11 datasets are also presented in Section \ref{techniqueAlgo} and Section \ref{data}, respectively.

On the basis of detailed analysis of results, key factors for the global adoption of online exams are identified and investigated. Particularly, in first step, the geographic distribution of participating countries in the selected studies is performed in Section \ref{researchCountry} as shown in Fig. \ref{fig1:Research_Outline}. Subsequently, the contributing countries are classified on the basis of developed and developing countries classes. Finally, the comparative analysis of adoptions factors with respect to key online exams features is performed in Section \ref{keyFactors}. The findings of this comparison facilitate countries and institutes for the selection of right online exam system on the basis of existing E-learning infrastructure and overall cost. To summarize, the identification and comparison of key factors in Section \ref{keyFactors} is a significant step towards the global adoption of online exams.

\section{Research Methodology}\label{methodology}
This research is conducted by utilizing the guidelines of Systematic Literature Review (SLR)~\cite{kitchenham2004procedures} where development of review protocol is an integral step. This section deals with the definition of primary categories (Section \ref{primarycategory}) and methodical design of review protocol (Section \ref{reviewProtocol}).
\subsection{Defining Categories}\label{primarycategory}
The selected studies are classified into three major categories (Section \ref{category}) in order to simplify the data extraction and synthesis process. The definitions of these categories are as follows:
\begin{enumerate}
  \item \textbf{Biometric Category:} The verification of examinee and prevention of cheating are two major challenges in online examination. In this regard, researchers frequently utilized biometric features like fingerprints, face, and head movements etc. in order to provide reliable solutions. For example, authors in~\cite{cote2016video} utilized head movements to analyze the abnormal behavior of examinee. All such studies where biometric features are utilized with some partial proof-of-concept implementation (without full tool development) are placed under Biometric Category. 
    \item \textbf{Software Applications Category:} There are studies where a complete software application for online exams are proposed for different purposes. For example, G. Frankl et al.~\cite{7-frankl2011secure} propose a complete software application named “Secure Exam Environment” for the execution of online exams. Similarly, in another study~\cite{rajala2016automatically}, ViLLE tool is developed to accomplish the automated assessment in online exams. All such studies where a complete tool is developed to achieve particular online exams objectives are placed under Software Applications Category.
  \item \textbf{General Category:} In few studies, biometric features along with other attributes (e.g. System calls, question bank generation etc.) are utilized to develop a full fledge online exam management system. For example, Moukhliss Ghizlane et al.~\cite{ghizlane2019new} developed a complete system comprising management and monitoring components where monitoring is accomplished through biometric features and other attributes are used for online exam management. Such studies simultaneously targeting both biometric and software application categories that are placed in General Category.  On the other hand, there are few studies (e.g.~\cite{subramanian2018using}) where conceptual framework is proposed. Furthermore, few studies proposed certain techniques (e.g. improved online exams user interface~\cite{karim2016proposed} etc.) which do not belong to either Biometric or Software Applications categories. Such multidisciplinary studies are also placed in the General category.
\end{enumerate}
    
 \subsection{Review Protocol}\label{reviewProtocol}
 The development of review protocol involves six steps as per the standard SLR guidelines~\cite{kitchenham2004procedures}. The first two steps (i.e. background and research questions) are already performed in the introduction (Section \ref{intro}) of the article. The details of the remaining four steps (i.e. Inclusion and exclusion criteria, Search process, Quality assessment and data extraction / synthesis) are given in subsequent sections.
 \subsubsection{Inclusion and Exclusion criteria}\label{InclusionExclusionCriteria}
 The inclusion and exclusion criteria are the most important part of SLR. Particularly, the studies are selected or rejected on the basis of this criteria. We develop 6 parameters for the inclusion and exclusion of studies as follows:
 \begin{enumerate}
  \item \textbf{Subject:} The selected study must belong to online exams substantially. Studies dealing with E-learning as a whole but discussing online exams marginally should be discarded.
  
\noindent \emph{Description:} In this SLR, online exams is a major subject. Therefore, the study should only be included where the improvement in online exams is the major concern. In fact, there exist studies (e.g.~\cite{andersson2018key}) where solution for different aspects of E-learning is proposed and online exams are discussed / researched marginally. Such studies are excluded as online exams is a major area of research for this SLR. 
\item \textbf{Application Research:} The study should only be selected if some genuine framework, technique, or software / prototype is proposed for the improvement of online exams. 

\noindent \emph{Description:} This SLR only considers studies that are dealing with application research. Particularly, only those studies are selected where some genuine technique, framework or software / prototype is proposed to improve certain aspect(s) of online exams. In this context, review studies (e.g.~\cite{boitshwarelo2017envisioning}) are not considered. Furthermore, empirical studies dealing with some particular hypothesis without any genuine proposal (e.g.~\cite{milone2017impact}) are also discarded. 
	
\item \textbf{Publication Year:} This SLR only considers studies, which are published from January 2016 to July 2020. 

\noindent \emph{Description:} Generally, latest studies are based on the findings / background of previous studies. Therefore, we have selected well-balanced publication year duration (i.e. January 2016 to July 2020) for this SLR. This duration not only covers latest online exams developments but also encompasses previous contributions logically. For example, latest features of Secure Exam Environment (SEE)~\cite{frankl2018pathways} for online exams are explored in 2017, however, its actual development was started in 2011~\cite{7-frankl2011secure}. Therefore, study~\cite{frankl2018pathways} is covering all aspects of SEE from 2011 to 2017.

\item \textbf{Publisher:} Six renowned scientific databases are considered in this SLR for the selection of studies as follows: 
\begin{itemize}
    \item IEEE
    \item Elsevier
    \item Springer
    \item ACM
    \item Wiley
    \item Taylor \& Francis
\end{itemize}

\noindent \emph{Description:} The studies are published in different scientific repositories. However, studies can be included without peer review in repositories like Google Scholar etc. On the other hand, the databases like IEEE, Springer, and Elsevier are trustworthy where studies are usually included in these repositories after peer review. Therefore, six most renowned scientific databases (i.e. IEEE, Elsevier, Springer, ACM, Wiley and Taylor \& Francis) are considered in this SLR for the selection of studies.

\item \textbf{Validation of Proposal:} The study should only be selected if the proposed approach is properly validated through suitable techniques like experimentation, prototyping etc.

\noindent \emph{Description:} The proper validation of proposal is really important for the high-quality research. Therefore, the study can only be selected if proper validation of proposal is performed through reliable techniques like experimentation, prototyping etc. In this context, there exist studies where the insufficient details are provided for validation of proposal. For example, Abisado et al.~\cite{abisado2019modeling} proposed machine learning approach for the analysis of online exams. However, all the details are summarized in one page and validation is discussed in only few lines. In another study~\cite{5-alrubaish2019automated} , the cheating prevention algorithm for online exams is proposed with sufficient details, however, the validation information is totally missing. Consequently, all such studies with insufficient / missing validation details are discarded during the SLR.

\item \textbf{Repetition:} Multiple studies having similar research contents are analyzed first and only one with most reliable contents is selected.

\noindent \emph{Description:} In the literature, there exist studies that present similar research contents. Particularly, the researchers usually propose the initial technique in some relevant conference. Subsequently, the full approach including complete implementation details are published in journal. For example, Ullah el al.~\cite{ullah2018multi} initially proposed dynamic profile questions approach for the authentication of online exams. Subsequently, authors published the extended version in~\cite{ullah2019dynamic} with complete details. In this SLR, we discard studies having almost similar research contents and we only select one of them with most reliable contents e.g. in aforementioned case, we select study~\cite{ullah2019dynamic}. 
\end{enumerate}
We performed this SLR on the basis of aforementioned inclusion and exclusion parameters. Particularly, the study is only selected if it completely follows all inclusion and exclusion parameters. The study is discarded even if a single inclusion and exclusion parameter is violated. 

\subsubsection{Search Process}\label{search}
We performed the search process through six databases (Section \ref{InclusionExclusionCriteria}) in order to select the relevant studies as per inclusion and exclusion criteria. The summary of search terms used in the search process is given in Table \ref{tab1:SearchProcess}. Particularly, we started the search process through most relevant search terms like “Online Exams” etc. However, such search terms returned thousands of results, which could not be fully analyzed. For example, Elsevier database (sciencedirect.com) returned 36,472 results in default settings for “Online Exams” search term. To optimize the search results, we utilized different filters like “Publication Year” (2016-2020), AND operator etc. to get the most relevant results. Similarly, we also used advance search options like “Where Title or Abstract Contains” etc. to speed up the search process. After applying several filters and advance search options, we were able to get optimum and most relevant results that could be completely analyzed. For example, we only got 131 search results regarding “Online Exams” search term from IEEE database after applying different filters e.g. publication year between 2016 to 2020 . Similarly, we got 46 results for “e-learning Exams” search term as given in Sr. \# 2 of Table \ref{tab1:SearchProcess}. In the same way, we applied different filters in each scientific repository and got the filtered results as given in third column of Table \ref{tab1:SearchProcess}.

Initially, we used simple and most relevant search terms like “Online Exams”, “e-learning Exams”, “Digital Exams” and “Electronic Exams” as given in Sr. \# 1 to 4 of Table \ref{tab1:SearchProcess}. Once we analyzed the search results of these simple terms, we found certain keywords that are frequently associated with online exams subject and could be utilized to find the relevant studies effectively. For example, we found that the terms like “Proctoring” and “Biometric” is frequently utilized while performing the authentication of examinees in online exams. Similarly, the terms like “Assessment” and “Question Bank” are frequently utilized in the context of online exams. Therefore, we developed more advanced and intelligent search terms like “Exam Proctoring”, “e-learning Assessment”, “Exams Biometric” and “Online Question Bank”, as given in Sr. 5 to 8 of Table \ref{tab1:SearchProcess}, in order to get relevant studies. These advanced search terms enabled us to find and select the studies, which cannot be picked through simple search terms.
\begin{table}
\caption{Search Process comprises different search terms and corresponding filtered results in each database}
\vspace*{-3mm}
\renewcommand{\arraystretch}{1.5} 
\linespread{0.95}\selectfont
\centering
\begin{adjustbox}{max width=\columnwidth}
       \begin{tabular}{clcccccc}\\\hline
        & &\multicolumn{6}{c}{Filtered Results}\\\cline{3-8}
       
       No. & Search Terms & IEEE & Elsevier &	Springer	& ACM &	Wiley &	Taylor \& Francis \\\hline\hline
         
1 &	Online Exams &	        131 & 130 &	156 &	28 & 6  & 14 \\ \cline{1-8}
2 &	E-learning Exams &	    46  & 2   &	4	&   76 & 18 & 10 \\ \cline{1-8}
3 &	Digital Exams &	        67  & 54  &	119	&   44 & 2  & 3 \\ \cline{1-8}
4 &	Electronic Exams &	    109 & 10  &	25	&   11 & 14 & 8 \\ \cline{1-8}
5 &	Exam Proctoring &	    13  & 14  &	19	&   2  & 1  & 3 \\ \cline{1-8}
6 &	E-learning Assessment &	265 & 7	  & 25	&   11 & 35 & 19 \\ \cline{1-8}
7 &	Exams Biometric &	    2   & 0   & 5 	&   8  & 2  & 0 \\ \cline{1-8}
8 &	Online Question Bank &	37  & 16  &	4	&   3  & 8  & 2 \\ \hline\hline
        \end{tabular}
\end{adjustbox}
    \vspace*{-5mm}
        \label{tab1:SearchProcess}
\end{table}

\begin{figure}[hbt!]
    \centering
    \includegraphics[width=\textwidth,]{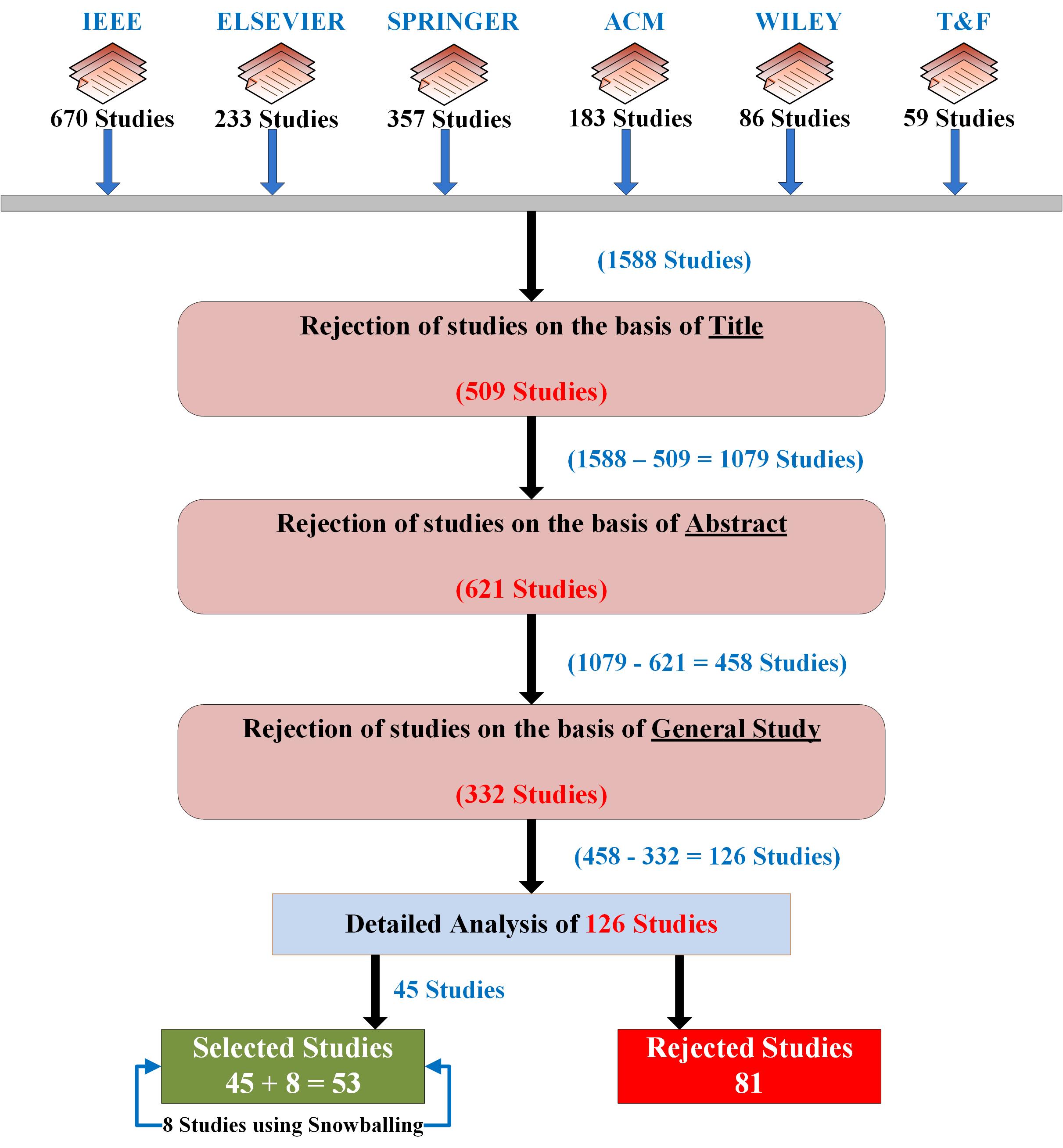}
    \caption{Search Process}
    \label{fig2:SearchProcess}
\end{figure}

The detailed investigation of search results was performed for the selection of studies as per inclusion and exclusion criteria. Particularly, the parameter \# 3 (Publication Year) and parameter \# 4 (Publisher) of inclusion and exclusion criteria were already ensured during search process. However, to guarantee the compliance of other parameters, the search results were systematically analyzed through different steps as shown in Fig. \ref{fig2:SearchProcess}.

\begin{itemize}
\item	Overall, we analyzed 1588 search results. Initially, we checked the titles of studies to confirm the relevance as per inclusion and exclusion criteria. The title of few studies clearly indicated their irrelevance with the given subject i.e. online exams. Therefore, we excluded 509 such studies by only analyzing the titles. 
\item	In the next step, the abstracts of remaining 1079 studies were investigated. It is observed that the abstracts of few studies clearly violating parameter \# 1 (Subject) and parameter \# 2 (Application Research) of inclusion and exclusion criteria. Therefore, we excluded 621 such studies by evaluating the contents of abstract.
\item	To this point, we had 458 remaining studies and most of them were following parameter \# 1 (Subject) and parameter \# 2 (Application Research) of inclusion and exclusion criteria. However, to confirm parameter \# 5 (Validation of Proposal) of inclusion and exclusion criteria, it was required to read different sections of those studies. Therefore, we  performed a general reading of remaining 458 studies where different sections were read summarily without going into in-depth details. As a result, we found 332 studies violating parameter \# 5, therefore, we excluded them as well.
\item	At this stage, we had 126 remaining studies. We performed  detailed analysis of these studies by reading each section carefully to ensure their compliance with all six parameters of inclusion and exclusion criteria. We found that few studies were still violating parameter \# 5. Furthermore, we also found repetition of few studies that was the violation of parameter \# 6 (Repetition). Therefore, we excluded 81 such studies and selected 45 remaining studies, which were fully compliant with all inclusion and exclusion parameters. 
\item	In the final stage, we executed snowballing process on 45 selected studies. The snowballing ensures the selection of relevant studies that have been accidentally missed during search process. For comprehensive exploration, both forward and backward snowballing techniques are utilized. As a result, we found 22 studies that seemed relevant in the given research context. After detailed analysis, we selected 8 studies, which were fully compliant with inclusion and exclusion criteria. Finally, 53 studies (45 + 8) were selected for further analysis in order to get the realistic and trustworthy answers to our research questions.    
\end{itemize}

\subsubsection{Quality Assessment}\label{quality}
We systematically developed the inclusion and exclusion criteria (Section \ref{InclusionExclusionCriteria}) that inherently supports the high-quality outcomes of this SLR. Particularly, the parameter 2 (Application Research) of inclusion and exclusion criteria ensures the selection of studies where some genuine technique, framework and software / prototype is proposed. This significantly improves the quality of this SLR as review articles and other insignificant studies are not considered. In addition to this, parameter 3 (publication years) ensures the selection of latest studies only. This leads to identify and analyze the current online exams developments. The year wise distribution of selected studies is given in Table \ref{tab2:yearWise}. It can be seen from the last column (Total) of Table \ref{tab2:yearWise} that we selected 8, 11, 15, 13 and 6 studies from 2016, 2017, 2018, 2019 and 2020, respectively. Therefore, 85\% selected studies are published during 2017-2020. The 64\% selected studies are published during last two and half years (i.e. Jan 2018 to July 2020). Consequently, the findings of this SLR are up-to-date due to the selection of latest studies. 

\begin{table}
\caption{Year wise distribution of selected studies}
\vspace*{-3mm}
\renewcommand{\arraystretch}{1.5} 
\linespread{0.95}\selectfont
\centering
\begin{adjustbox}{max width=\columnwidth}
       \begin{tabular}{cc p{10cm} c}\\\hline
    
       No. & Year & Studies & Total \\\hline\hline

1 &	2016 &	\cite{natawiguna2016virtualization},\cite{ketui2016item},\cite{cote2016video},\cite{prathish2016intelligent},\cite{mahatme2016data},\cite{fan2016gesture},\cite{rajala2016automatically},\cite{karim2016proposed} &	8 \\ \cline{1-4}
2 &	2017 &	\cite{atoum2017automated},\cite{mathapati2017secure},\cite{shi2017research},\cite{traore2017ensuring},\cite{sabbah2017security},\cite{ettarres2017evaluation},\cite{kassem2017formal},\cite{kar2017novel},\cite{diedenhofen2017pagefocus},\cite{d2017conceptual},\cite{chuang2017detecting} &	11 \\ \cline{1-4}
\multirow{2}{*}{3} & \multirow{2}{*}{2018} &	\cite{aisyah2018development},\cite{sukadarmika2018introducing},\cite{hu2018research},\cite{lemantara2018prototype},\cite{abisado2018towards},\cite{chen2018application},\cite{opgen2018application},\cite{zhang2018analysis}, &	\multirow{2}{*}{15}\\
& & \cite{subramanian2018using},\cite{frankl2018pathways},\cite{kolhar2018online},\cite{nandini2020automatic},\cite{albastroiu2018exam},\cite{ullah2019dynamic},\cite{baykasouglu2018process} & \\ \cline{1-4}
4 &	2019 &	\cite{asep2019design},\cite{ghizlane2019new},\cite{sukmandhani2019face},\cite{vomvyras2019exam},\cite{chua2019online},\cite{das2019examination},\cite{boussakuk2019online},\cite{wagstaff2019automatic},\cite{sultan2019automatically},\cite{fanani2019interactive},\cite{tashu2019intelligent},\cite{manoharan2019cheat},\cite{al2019integrated} &	13 \\ \cline{1-4} 
5 &	2020 &	\cite{garg2020convolutional},\cite{matveev2020virtual},\cite{jiang2019design},\cite{kausar2020fog},\cite{golden2020addressing},\cite{wu2020exam} &	6 \\ 
\hline\hline
        \end{tabular}
\end{adjustbox}
    \vspace*{-5mm}
        \label{tab2:yearWise}
\end{table}

The six most renowned and trustworthy scientific databases are considered for the selection of studies as per parameter 4 of inclusion and exclusion criteria (Section \ref{InclusionExclusionCriteria}). This significantly improves the quality of this SLR. The distribution of selected studies with respect to scientific databases is given in Table \ref{tab3:SummaryPublisher}. It can be seen from Table \ref{tab3:SummaryPublisher} that 22 studies are selected from IEEE and 15 studies are selected from Springer. Furthermore, 8 studies are selected from ACM, 4 from Elsevier, 1 from Taylor \& Francis, and 3 are selected from Wiley. It is important to note that the selection of studies does not involve any sort of biasness. In fact, we properly searched and analyzed results from each database,   where relatively higher number of relevant studies were obtained from IEEE and Springer. In contrast, fewer number of studies were extracted from Elsevier, Wiley and Taylor \& Francis.

\begin{table}
\caption{Summary of Selected Studies with respect to Publisher}
\vspace*{-3mm}
\renewcommand{\arraystretch}{1.5} 
\linespread{0.95}\selectfont
\centering
\begin{adjustbox}{max width=\columnwidth}
       \begin{tabular}{cllc}\\\hline
    
       No. & Scientific Database  & Studies & Total \\\hline\hline
\multirow{4}{*}{1} &	\multirow{4}{*}{IEEE} &	\cite{asep2019design},\cite{aisyah2018development},\cite{natawiguna2016virtualization},\cite{ghizlane2019new},\cite{ketui2016item},\cite{atoum2017automated}, & \multirow{4}{*}{22} \\
& & \cite{sukmandhani2019face}, \cite{cote2016video},\cite{sukadarmika2018introducing},\cite{prathish2016intelligent}, \cite{hu2018research}, & \\

& & \cite{garg2020convolutional},\cite{vomvyras2019exam},\cite{matveev2020virtual},\cite{chua2019online},\cite{mathapati2017secure}, \cite{shi2017research}, & \\

& & \cite{mahatme2016data},\cite{fan2016gesture},\cite{das2019examination},\cite{boussakuk2019online},\cite{lemantara2018prototype} & 
\\\cline{1-4}

2 &	ELSIVER &	\cite{golden2020addressing},\cite{karim2016proposed},\cite{wu2020exam},\cite{manoharan2019cheat} &	4 \\\cline{1-4}
\multirow{3}{*}{3} &	\multirow{3}{*}{SPRINGER} &	\cite{traore2017ensuring},\cite{subramanian2018using},\cite{frankl2018pathways},\cite{sabbah2017security},\cite{ettarres2017evaluation},\cite{tashu2019intelligent}, \cite{kolhar2018online}, & \multirow{3}{*}{15} \\ 
& &  \cite{kassem2017formal},\cite{jiang2019design},\cite{nandini2020automatic},\cite{albastroiu2018exam},\cite{kausar2020fog}, & \\

& & \cite{kar2017novel},\cite{ullah2019dynamic},\cite{diedenhofen2017pagefocus} &
\\\cline{1-4}
\multirow{2}{*}{4} & \multirow{2}{*}{ACM} &	\cite{abisado2018towards},\cite{chen2018application},\cite{wagstaff2019automatic},\cite{rajala2016automatically},\cite{opgen2018application}, &	\multirow{3}{*}{8} \\
& & \cite{zhang2018analysis},\cite{sultan2019automatically},\cite{fanani2019interactive} & \\
\cline{1-4}

5 &	WILEY &	\cite{d2017conceptual},\cite{al2019integrated},\cite{baykasouglu2018process} &	3 \\\cline{1-4}
6 &	Taylor \& Francis &	\cite{chuang2017detecting} &	1 \\
\hline\hline
        \end{tabular}
\end{adjustbox}
    \vspace*{-5mm}
        \label{tab3:SummaryPublisher}
\end{table}

\begin{figure}[hbt!]
    \centering
    \includegraphics[]{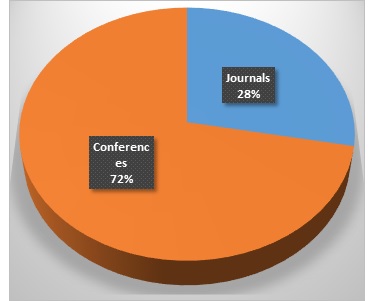}
    \caption{Distribution of Selected Studies with respect to publication type}
    \label{fig3:StudiesDistribution}
\end{figure}

The type (i.e. Journal or Conference) of selected study is another important factor for ensuring the quality of SLR. Although, we tried to select journal studies as much as possible, we were able to find 15 journal studies (out of 53), which were fully compliant with the inclusion and exclusion criteria. The distribution of selected studies on the basis of type is shown in Fig. \ref{fig3:StudiesDistribution} where 28\% of selected studies are from reputed journals while remaining 72\% studies are from conferences. It is important to note that two book chapters (i.e.~\cite{traore2017ensuring,sabbah2017security}) are selected from springer, however, we placed these book chapters in conference studies to keep the discussion simple.

\subsubsection{Data Extraction and Synthesis}\label{dataextraction}
The data extraction and synthesis process are executed after the selection of studies as per inclusion and exclusion criteria (Section \ref{InclusionExclusionCriteria}). This leads to find the realistic answers to research questions as defined in Section \ref{intro}. The data extraction template is defined, as given in Table \ref{tab4:dataExtractionElements}, to extract and analyze the elements of concerns from selected studies. Particularly, the primary elements are first extracted from selected studies as given from Sr. \# 1 to 4 of Table \ref{tab4:dataExtractionElements}. Subsequently, the data extraction with synthesis is performed to extract relevant elements, which are essential to answer RQs. For example, categorization of studies is performed to give the answer of RQ1. Similarly, the data extraction with synthesis is carried out for other important elements as given from Sr. \# 5 to 12 of Table \ref{tab4:dataExtractionElements}.

\begin{table*}
\caption{Elements of data extraction and synthesis}
\vspace*{-3mm}
\renewcommand{\arraystretch}{1.5} 
\linespread{0.95}\selectfont
\centering
\begin{adjustbox}{max width=\textwidth}
\begin{tabularx}{\linewidth}{c L p{6cm}}
\\\hline
\textbf{No.} & \textbf{Extracted Element}  & \textbf{Particulars} \\\hline\hline
1 &	Bibliographic information &	Title, Authors, Publisher and Publication year \\\cline{1-3}
2 &	Summary & The primary proposal of each selected study \\\cline{1-3}
3 &	Limitations & Limitations / assumptions (if any) of study while achieving particular objectives \\\cline{1-3}    
4 &	Validation method &	Validation method (e.g. case study, dataset etc.) used in each selected study \\\cline{1-3}

\multicolumn{3}{c}{\textbf{Data Extraction with Synthesis}}\\\cline{1-3}

5 &	Categorization & The classification of selected studies as per defined categories (Section \ref{primarycategory}). The results are summarized in Section \ref{category} (Table \ref{tab5:primarycategory}) \\\cline{1-3}
6 &	Targeted Online Exams Attributes & Leading online exam attributes targeted in selected studies. The results are summarized in Section \ref{leadingAttribute} (Table \ref{tab6:leadingExamFeatures}). \\\cline{1-3}
7 &	Underlying Development Approach & Underlying development approach used in each selected study for the implementation of a particular solution. The results are summarized in Section \ref{developmentApproach} (Table \ref{tab7:developmentApproaches}). \\\cline{1-3}
8 &	Techniques / Algorithms & Techniques / algorithms proposed in each study to achieve required objective. The results are summarized in Section \ref{techniqueAlgo} (Table \ref{tab8:LeadingTechniques}). \\\cline{1-3}
10 & Tools & Leading tools proposed and utilized in selected studies. The results are summarized in Section \ref{tool} (Table \ref{tab9:LeadingTools}) and (Table \ref{tab10:ExistingTools}). \\\cline{1-3}
11 &	Datasets & The important datasets used / proposed in selected studies. The results are summarized in Section \ref{data} (Table \ref{tab11:leadingDatasets}). \\\cline{1-3}
12 & Country of Research & Countries / institutions participated in selected study. The results are summarized in Section \ref{researchCountry} (Table \ref{tab12:ResearchCountry}). \\
\hline\hline
\end{tabularx}
\end{adjustbox}
    \vspace*{-5mm}
        \label{tab4:dataExtractionElements}
\end{table*}
To this point, we have discussed all major components of review protocol. Moreover, 53 studies have been selected on the basis of inclusion and exclusion criteria. Furthermore, data extraction is performed to extract the relevant information from selected studies. This leads to compile the results precisely as given in subsequent section. 
\section{Results}\label{result}
This section presents precise results to provide authentic answers to research questions. Particularly, the primary categorization of studies is performed in Section \ref{category}. Moreover, the leading online exams attributes and underlying development approaches are identified in Sections \ref{leadingAttribute} and \ref{developmentApproach}, respectively. Furthermore, the proposed techniques / algorithms are presented in Section \ref{techniqueAlgo}. In addition to this, the leading tools used and proposed by researchers are given in Section \ref{tool}. Finally, the important datasets and summary of countries participation in online exams research are provided in Section \ref{data} and Section \ref{researchCountry}, respectively.
\subsection{Categorization}\label{category}
The three main categories are defined in Section \ref{primarycategory} to perform the primary classification of selected studies. These categories are: 1) Biometric 2) Software Applications and 3) General. The categorization results are given in Table \ref{tab5:primarycategory}. It can be seen from Table \ref{tab5:primarycategory} that 7 and 11 studies belong to Biometric and Software Applications categories, respectively whereas 35 studies belong to General category. It is already expected that General category includes higher number of studies due to its multidisciplinary definition (Section \ref{primarycategory}). For example, the studies (e.g.~\cite{aisyah2018development,ghizlane2019new,atoum2017automated,sukmandhani2019face} etc.] simultaneously dealing with Biometric and Software Application categories are placed in General category. Moreover, the studies dealing with unique aspects (e.g. Virtualization~\cite{natawiguna2016virtualization} etc.) are also placed in General category. Similarly, studies dealing with the proposal of conceptual framework (e.g.~\cite{subramanian2018using} etc.) are also placed under this category. That is why, larger number of selected studies fall under General category in this study.

The seven studies included in Biometric category do not develop substantial software applications. Particularly, Biometric studies only propose particular technique / algorithm where a trivial proof-of-concept implementation is performed for validation. For example, Z. Fan et al.~\cite{fan2016gesture} proposed a novel approach to identify misbehavior of examinee through gesture movements using Kinect device. Although authors validated the proposed approach by implementing two modules, the proper software application is not developed. Therefore, we do not place such studies in General category since only Biometric approach is proposed without developing any proper software application. On the other hand, the studies in Application Software categories developed a complete tool as part of their research. For instance, M. Boussakuk et al.~\cite{boussakuk2019online} developed online assessment system named “CleverTesting”. It is important to mention that Software Applications category do not include any study where biometric characteristics are used for the development of tool. In fact, studies dealing with software tool using biometric features (e.g.~\cite{aisyah2018development}) are placed in General category.

\begin{table}[ht]
\caption{Classification of selected studies with respect to Primary categories}
\vspace*{-3mm}
\renewcommand{\arraystretch}{1.5} 
\linespread{0.95}\selectfont
\centering
\begin{adjustbox}{max width=\columnwidth}
       \begin{tabular}{cllc}\\\hline
    
No. & Category  & Reference & Total \\\hline\hline
1 &	Biometric &	\cite{asep2019design},\cite{cote2016video},\cite{hu2018research},\cite{garg2020convolutional},\cite{fan2016gesture},\cite{abisado2018towards},\cite{albastroiu2018exam} &	7 \\\cline{1-4}
\multirow{2}{*}{2} &	\multirow{2}{*}{Software Applications} &	\cite{ketui2016item},\cite{vomvyras2019exam},\cite{mathapati2017secure},\cite{boussakuk2019online},\cite{lemantara2018prototype},\cite{rajala2016automatically}, &	\multirow{2}{*}{11} \\

& & \cite{sultan2019automatically},\cite{frankl2018pathways},\cite{jiang2019design},\cite{kausar2020fog},\cite{al2019integrated} & \\
\cline{1-4}

\multirow{5}{*}{3} & \multirow{5}{*}{General}  &	\cite{aisyah2018development},\cite{natawiguna2016virtualization},\cite{ghizlane2019new},\cite{atoum2017automated},\cite{sukmandhani2019face},\cite{sukadarmika2018introducing}, &\multirow{5}{*}{35}\\
& & \cite{prathish2016intelligent},\cite{matveev2020virtual},\cite{chua2019online}, \cite{shi2017research}, \cite{mahatme2016data}, \cite{das2019examination},\cite{chen2018application},\cite{wagstaff2019automatic}, & \\
& & \cite{opgen2018application},\cite{zhang2018analysis},\cite{fanani2019interactive},\cite{traore2017ensuring}, \cite{subramanian2018using},\cite{sabbah2017security}, \cite{ettarres2017evaluation}, \cite{tashu2019intelligent}, & \\
& &  \cite{kolhar2018online},\cite{kassem2017formal},\cite{nandini2020automatic},\cite{kar2017novel},\cite{ullah2019dynamic},\cite{diedenhofen2017pagefocus},\cite{golden2020addressing}, & \\

& & \cite{karim2016proposed},\cite{wu2020exam},\cite{manoharan2019cheat},\cite{d2017conceptual},\cite{baykasouglu2018process},\cite{chuang2017detecting} & \\

\hline\hline
        \end{tabular}
\end{adjustbox}
    \vspace*{-5mm}
        \label{tab5:primarycategory}
\end{table}

\subsection{Leading Online Exams Attributes}\label{leadingAttribute}
There are several associated attributes while executing the online exams. For example, one important attribute is the verification / authentication of examinee during online exams. In order to analyze the targeted online exams attributes in selected studies, following leading attributes classes are defined:  
\begin{enumerate}
\item \textbf{Verification \& Abnormal Behavior:} The verification \& abnormal behavior detection of examinee are highly important features in online exams. There are two types of verification~\cite{ghizlane2019new} i.e. static and continuous. In static verification, examinee is verified only once at the beginning of online exam. In continuous verification, the authentication / verification of examinee is continuously performed after certain time periods during the online exam. Likewise, prevention of cheating through detection of abnormal behavior is very important to ensure the fairness in online exams. The verification of examinee and detection of abnormal behavior are closely related concepts. For example, the biometric characteristics are frequently utilized for both verification and detection of abnormal behavior of examinee. Therefore, the studies dealing with examinee verification and / or detection of abnormal behavior are placed under \emph{Verification / Abnormal Behavior} class. 
\item \textbf{Security:} The security of online exams is an important feature where unauthorized access to different system components (e.g. user management, questions bank etc.) is assured. The studies dealing with different aspects of security in online exams are placed under \emph{Security} class.
\item \textbf{Question Bank Generation \& Evaluation:} The automatic generation of question bank and evaluation of examinee answers are highly important in online exams. There are studies (e.g.~\cite{wu2020exam}) where certain techniques are proposed for the generation of multiple-choice and / or descriptive questions for online exams. Moreover, there are studies (e.g.~\cite{vomvyras2019exam}) where automatic evaluation of examinee answers is performed. All such studies are placed under \emph{Question Bank Generation \& Evaluation} class. 
\item \textbf{Usability:} The simplicity and user-friendliness are also important characteristics of online exam systems. There exist studies (e.g.~\cite{karim2016proposed}) to improve the user interface design of online exams systems. All such studies are placed under \emph{Usability} class. 
\item \textbf{Other:} The studies simultaneously dealing with more than one of the aforementioned features classes are placed in \emph{Other} class. For example, both \emph{Verification / Abnormal Behavior} and \emph{Security} classes are targeted simultaneously in the study~\cite{ghizlane2019new}.

\end{enumerate}
The aforementioned online exams feature classes, targeted in the selected studies, are given in the Table \ref{tab6:leadingExamFeatures}. It can be seen from the Table \ref{tab6:leadingExamFeatures} that Question Bank Generation \& Evaluation (20 studies) and Verification \& Abnormal Behavior (19 Studies) are the most frequently targeted features in the selected studies. Particularly, the researchers commonly tried to improve the question bank generation and automatic assessment of answers in online exams. For example, Zhengyang Wu et al.~\cite{wu2020exam} proposed an AI approach for the effective generation of question bank in order to improve the overall assessment in online exams. In another study~\cite{lemantara2018prototype}, authors proposed a novel approach for the detection of plagiarism in online exams to automatically evaluate the answers swiftly.  Similarly, Verification \& Abnormal Behavior feature class is also an attractive area for researchers where several techniques have been proposed to ensure the integrity of online exams. For example, Diedenhofen et al.~\cite{diedenhofen2017pagefocus} developed PageFocus JavaScript to assess the abnormal events in examinee’s system for cheating prevention. Similarly, there are studies (e.g.~\cite{asep2019design,aisyah2018development}) where the verification / authentication of examinee is ensured.

We identified 9 studies to be categorized in the Other class where different feature classes are targeted simultaneously. For example, Ghizlane et al.~\cite{ghizlane2019new} proposed an approach for continuous monitoring of online exam where examinee authentication is performed through face recognition technique. A security model is also proposed to ensure communication between sever and clients remains secure. In this way, the authors are simultaneously targeting Verification \& Abnormal Behavior as well as Security classes. Similar is the case with study~\cite{sabbah2017security}.

\begin{table}[ht]
\caption{Leading online exam features targeted in the selected studies }
\vspace*{-3mm}
\renewcommand{\arraystretch}{1.5} 
\linespread{0.95}\selectfont
\centering
\begin{adjustbox}{max width=\columnwidth}
       \begin{tabular}{cllc}\\\hline
    
No. & Category  & Reference & Total \\\hline\hline
\multirow{3}{*}{1} & \multirow{2}{*}{Verification \&} &
\cite{asep2019design},\cite{aisyah2018development},\cite{atoum2017automated},\cite{sukmandhani2019face},\cite{cote2016video}, &	\multirow{3}{*}{19}
 \\
&	\multirow{2}{*}{Abnormal Behavior} & \cite{prathish2016intelligent},\cite{hu2018research},\cite{garg2020convolutional},\cite{fan2016gesture},\cite{abisado2018towards}, \cite{opgen2018application},\cite{traore2017ensuring},	& \\

& & \cite{subramanian2018using},\cite{kolhar2018online},\cite{kassem2017formal},\cite{albastroiu2018exam},\cite{ullah2019dynamic},\cite{diedenhofen2017pagefocus},\cite{chuang2017detecting} & \\
 \cline{1-4}

2 &	Security &	\cite{mathapati2017secure},\cite{kausar2020fog},\cite{sukadarmika2018introducing} &	3 
 \\\cline{1-4}

\multirow{3}{*}{3} &	\multirow{2}{*}{Question Bank Generation} &	\cite{ketui2016item},\cite{vomvyras2019exam},\cite{matveev2020virtual},\cite{mahatme2016data},\cite{das2019examination},\cite{boussakuk2019online}, &	\multirow{3}{*}{20}
 \\
 &	\multirow{2}{*}{\& Evaluation} & \cite{lemantara2018prototype},\cite{chen2018application},\cite{wagstaff2019automatic},\cite{rajala2016automatically}, \cite{zhang2018analysis},\cite{sultan2019automatically},\cite{ettarres2017evaluation},	&	\\
 
 & & \cite{tashu2019intelligent},\cite{jiang2019design},\cite{nandini2020automatic},\cite{kar2017novel},\cite{golden2020addressing},\cite{wu2020exam},\cite{manoharan2019cheat} & \\
 
 \cline{1-4}

4 &	Usability &	\cite{fanani2019interactive},\cite{karim2016proposed} & 2 \\\cline{1-4}

\multirow{2}{*}{5} &	\multirow{2}{*}{Other} &	\cite{natawiguna2016virtualization},\cite{ghizlane2019new},\cite{chua2019online},\cite{shi2017research},\cite{frankl2018pathways},\cite{sabbah2017security}, &	\multirow{2}{*}{9}\\

& & \cite{d2017conceptual},\cite{al2019integrated},\cite{baykasouglu2018process} & \\
\hline\hline
        \end{tabular}
\end{adjustbox}
    \vspace*{-5mm}
        \label{tab6:leadingExamFeatures}
\end{table}

On this basis of Table \ref{tab6:leadingExamFeatures} statistics, it can be concluded that Question Bank Generation \& Evaluation is the most frequently researched feature class during the past five years followed by Verification \& Abnormal Behavior class. On the other hand, usability feature class is least targeted by researchers during past five years followed by security feature class.

\subsection{Underlying Development Approaches}\label{developmentApproach}
So far, we have developed primary categories (Section \ref{category}) to classify selected studies on the basis of general areas. Additionally, feature based classes are also developed (Section \ref{leadingAttribute}) to organize the selected studies on the basis of important features of online exams. However, to answer RQ3, it is required to group the selected studies on the basis of underlying development approaches. To achieve this, following categories are defined on the basis of important development approaches: 
\begin{enumerate}
    \item \textbf{Machine Learning:} There are studies where certain Machine Learning (ML) concepts such as feature selection, classification etc. are utilized to propose particular approach / technique for online exams. All such studies are placed in \emph{Machine Learning} category. 
    \item \textbf{Artificial Intelligence:} Although machine learning and Artificial Intelligence (AI) are highly overlapping theories, here we distinguish both through certain concepts in the given research context. Particularly, the studies dealing with Natural Language Processing (NLP), dynamic programming and genetic algorithms are placed under \emph{Artificial Intelligence} category.
    \item \textbf{Formal Methods:} The Formal Methods (FMs) such as z-notations, timed automata etc. are frequently utilized for system development in different domains like embedded systems~\cite{anwar2020unified} etc. In the context of online exams, it is interesting to investigate the application of FMs. Therefore, the studies using formal methods to propose some novel online exams solution are placed under \emph{Formal Methods} category.
    \item \textbf{Traditional Development:} There are studies where different programming languages like Java, C\#, PHP etc. are utilized to develop some desktop / web-based solution for online exams. It is important to note that such studies do not utilize ML, AI or FMs techniques for system development. All such studies are placed under \emph{Traditional Development} category. 
    \item \textbf{Additional:} In few studies (e.g.~\cite{sukmandhani2019face}), a complete online exam solution with advanced features is provided by utilizing both ML / AI and traditional development methods. Furthermore, there are studies where conceptual frameworks and other techniques are proposed, which are not relevant to AI, ML, FMs or traditional development categories. All such studies are placed under \emph{Additional} category.

\end{enumerate}

The summary of underlying development approaches in selected studies is given in Table \ref{tab7:developmentApproaches}.  We identified 11 and 9 studies where proposal is based on some ML and AI techniques, respectively. For example, Nandini and Maheswari ~\cite{nandini2020automatic} proposed ML based technique to evaluate the answers to descriptive questions automatically in online exams. Particularly, syntactical approach is introduced for feature extraction whereas classification is performed using naïve bayes. In another study~\cite{mahatme2016data}, an AI approach is introduced by combining data mining with fuzzy logic concepts for the intelligent categorization of question bank in online exams. On the other hand, we identified 15 studies that utilized traditional development languages such as Java, PHP etc.  without employing AI or ML approaches. For example, S. Kausar et al.~\cite{kausar2020fog} proposed secure communication mechanism in online exams by utilizing the concepts of fog computing. Authors did not utilize any ML, AI or FMs technique and proof of concept implementation is accomplished through traditional development platform i.e. Asp .NET and C\#. Similarly, in another study~\cite{jiang2019design}, the authors developed a web-based online exam system using PHP programming language.

It can be seen from Table \ref{tab7:developmentApproaches} that there are 17 studies in \emph{Additional} category where ML and AI techniques are combined with traditional development to propose a complete online exam solution for particular purposes. For example, Sabbah Y.W.~\cite{sabbah2017security} proposed a complete online exams framework and tool with two major components i.e. Interactive and Secure e-Examination Unit (ISEEU) and Smart Approach for Bimodal Biometrics Authentication in Home-exams (SABBAH). Author proposed AI-based algorithms for system development and then integrated with Moodle~\cite{84moodle} using PHP. In addition to such studies, \emph{Additional} group also contains few studies, which cannot be placed in any other group due to their unique characteristics. For example, Ullah et al.~\cite{ullah2019dynamic} proposed a unique cheating prevention approach in online exams by utilizing the concept of dynamic profile questions. Authors designed several questions, without employing any ML / AI technique or traditional development, in order to improve the user authentication in online exams for cheating prevention.

\begin{table}[ht]
\caption{Underlying development approaches employed in the selected studies}
\vspace*{-3mm}
\renewcommand{\arraystretch}{1.5} 
\linespread{0.95}\selectfont
\centering
\begin{adjustbox}{max width=\columnwidth}
       \begin{tabular}{cllc}\\\hline
    
       No. & Category  & Reference & Total \\\hline\hline
\multirow{2}{*}{1} &	\multirow{2}{*}{Machine Learning}  & 	\cite{asep2019design},\cite{cote2016video},\cite{hu2018research},\cite{garg2020convolutional},\cite{abisado2018towards},\cite{chen2018application}, &	\multirow{2}{*}{11} \\

& & \cite{wagstaff2019automatic},\cite{opgen2018application},\cite{nandini2020automatic},\cite{albastroiu2018exam},\cite{chuang2017detecting} & \\

\cline{1-4}

\multirow{2}{*}{2} &	\multirow{2}{*}{Artificial Intelligence} & 	\cite{matveev2020virtual},\cite{mahatme2016data},\cite{fan2016gesture},\cite{das2019examination},\cite{lemantara2018prototype},\cite{zhang2018analysis}, &	\multirow{2}{*}{9}\\
& & \cite{kar2017novel},\cite{wu2020exam},\cite{baykasouglu2018process} & \\
\cline{1-4}
3 &	Formal Methods &	\cite{kassem2017formal} &	1 \\\cline{1-4}

\multirow{3}{*}{4} & \multirow{3}{*}{Traditional Development} &	\cite{aisyah2018development},\cite{ketui2016item},\cite{vomvyras2019exam},\cite{chua2019online},\cite{mathapati2017secure},\cite{shi2017research},&	\multirow{3}{*}{15} \\
& & \cite{boussakuk2019online},\cite{rajala2016automatically},\cite{sultan2019automatically},\cite{frankl2018pathways},\cite{kolhar2018online},  & \\

& & \cite{jiang2019design},\cite{kausar2020fog},\cite{diedenhofen2017pagefocus},\cite{manoharan2019cheat} &

\\\cline{1-4}
\multirow{3}{*}{5} &	\multirow{3}{*}{Additional} &	\cite{natawiguna2016virtualization},\cite{ghizlane2019new},\cite{atoum2017automated},\cite{sukmandhani2019face},\cite{sukadarmika2018introducing}, &	\multirow{3}{*}{17} \\
& & \cite{prathish2016intelligent}, \cite{fanani2019interactive},\cite{traore2017ensuring},\cite{subramanian2018using},\cite{sabbah2017security},\cite{ettarres2017evaluation},\cite{tashu2019intelligent}, & \\
& & \cite{ullah2019dynamic},\cite{golden2020addressing},\cite{karim2016proposed},\cite{d2017conceptual},\cite{al2019integrated} & \\
\hline\hline
        \end{tabular}
\end{adjustbox}
    \vspace*{-5mm}
        \label{tab7:developmentApproaches}
\end{table}

\subsection{Techniques / Algorithms}\label{techniqueAlgo}
In the selected studies, several techniques / algorithms have been proposed to achieve a particular objective for the improvement of online exams. The summary of leading techniques / algorithms proposed in the selected studies is given in Table \ref{tab8:LeadingTechniques}. Researchers proposed ML techniques, based on CNN, for examinee verification~\cite{asep2019design}, cheating prevention~\cite{hu2018research,garg2020convolutional}and online exams based techniques to improve verification / abnormal behavior feature (i.e.~\cite{asep2019design,hu2018research,garg2020convolutional} and automatic assessment~\cite{wagstaff2019automatic}. Similarly, researchers proposed different techniques / algorithms for face recognition and head pose estimation / detection as given in serial \# 2 and 3 of Table \ref{tab8:LeadingTechniques}, respectively. Furthermore, different NLP based techniques and genetic algorithms are proposed as given in serial \# 4 and 5 of Table \ref{tab8:LeadingTechniques}, respectively. In addition to this, few researchers proposed highly unique techniques to improve certain aspects of online exams. For example, Kassem et al.~\cite{kassem2017formal} proposed a FMs based approach, using Quantified Event Automata and $\pi$-calculus, to assess the violations in online exams. 

It is important to note that we only present leading techniques / algorithms in Table \ref{tab8:LeadingTechniques} and trivial proposals are not included for simplicity. For example, Abisado et al.~\cite{abisado2018towards} proposed simple divide and conquer algorithm for the detection of abnormal behavior during online exams. In another study~\cite{chuang2017detecting}, standard logistic regression model without any significant variation is used to predict cheating in online exams. Therefore, we do not include such trivial techniques in Table \ref{tab8:LeadingTechniques}. It is important to mention that proper information regarding proposed technique / algorithm is not available in some of the studies. For example, S. Aisyah et al.~\cite{aisyah2018development} developed online exams authentication system with two components i.e. authentication and supervision. However, authors did not provide any substantial information about underlying techniques / algorithms employed for the system development. Therefore, such studies are not included in Table \ref{tab8:LeadingTechniques}.

\begin{table}[ht]
\caption{Leading Techniques / Algorithms Proposed in the Selected Studies }
\vspace*{-3mm}
\renewcommand{\arraystretch}{1.5} 
\linespread{0.95}\selectfont
\centering
\begin{adjustbox}{max width=\columnwidth}
       \begin{tabular}{clr}\\\hline
    
       No. & Techniques / Algorithms & Reference  \\\hline\hline
1 &	Convolutional Neural Networks (CNN) Based Techniques &	\cite{asep2019design},\cite{hu2018research},\cite{garg2020convolutional},\cite{wagstaff2019automatic} \\\cline{1-3}
2 &	Face Recognition Techniques / Algorithms &	\cite{ghizlane2019new},\cite{atoum2017automated},\cite{sukmandhani2019face},\cite{traore2017ensuring} \\\cline{1-3}
3 &	Head Pose Estimation and Detection Techniques & \cite{cote2016video},\cite{fanani2019interactive},\cite{chuang2017detecting} \\\cline{1-3}
4 &	Natural Language Processing (NLP) Based Techniques  &	\cite{matveev2020virtual},\cite{das2019examination},\cite{kar2017novel} \\\cline{1-3}
5 &	Genetic Algorithms &	\cite{zhang2018analysis},\cite{jiang2019design},\cite{wu2020exam} \\\cline{1-3}
6 &	Rule Based Inference Technique &	\cite{prathish2016intelligent},\cite{hu2018research} \\\cline{1-3} 
7 &	Semantic Similarity Technique &	\cite{tashu2019intelligent},\cite{baykasouglu2018process} \\\cline{1-3}
8 &	Hardware / Software Virtualization Technique &	\cite{natawiguna2016virtualization} \\\cline{1-3}
9 &	Time Adaptive for Mobile E-Exam (TAMEx) Technique &	\cite{sukadarmika2018introducing} \\\cline{1-3}
10 &	Fuzzy Clustering Technique  &	\cite{mahatme2016data} \\\cline{1-3}
11 &	Two dimensional Gesture Detection Technique & 	\cite{fan2016gesture} \\\cline{1-3}
12 &	K Means Clustering and Rule Mining Algorithms &	\cite{chen2018application} \\\cline{1-3}
13 &	Bayesian Network based Technique  &	\cite{ettarres2017evaluation} \\\cline{1-3}
14 &	Quantified Event Automata  and $\pi$-calculus Technique &	\cite{kassem2017formal} \\\cline{1-3}
15 &	Dynamic Profile Questions Technique &	\cite{ullah2019dynamic} \\\cline{1-3}
16 &	Syntactical Relational Feature Extraction Technique &	\cite{nandini2020automatic}

 \\ \hline\hline
        \end{tabular}
\end{adjustbox}
    \vspace*{-5mm}
        \label{tab8:LeadingTechniques}
\end{table}
\subsection{Tools}\label{tool}
This section presents the tools that have been proposed as well as utilized in the selected studies. The tools developed/ implemented as a part of research in the selected studies are given in Section \ref{proposedtool}. Likewise, existing tools used in the selected studies for the implementation of proposed technique / tool are given in Section \ref{utilizedtool}.

\subsubsection{Proposed Tools}\label{proposedtool}
Development of a tool is an important aspect for the improvement of online exams. In the selected studies, several tools have been developed supporting various features of online exams. The summary of proposed / developed tools in the selected studies is provided in Table \ref{tab9:LeadingTools}. The tool name is given in second column of Table \ref{tab9:LeadingTools}. The third column highlighted the online exams features, which are supported by the given tool. The fourth column identifies the availability of a given tool i.e. Public, Proprietary and Not-Applicable (N-A). The proposed tool where web link or address is provided for download is identified as Public whereas the tool, which requires licensing is recognized as Proprietary. The tools where availability related information (e.g. web link etc.) is not provided are identified as Not-Applicable (N-A). References of the studies proposing the given tools are provided in the last column.

\begin{table*}
\caption{Leading Online Exams Tools Proposed in the Selected Studies}
\vspace*{-3mm}
\renewcommand{\arraystretch}{1.5} 
\linespread{0.95}\selectfont
\centering
\begin{adjustbox}{max width=\textwidth}
       \begin{tabular}{clccccr}\\\hline
        
       & &\multicolumn{3}{c}{Supported Features} & & \\\cline{3-5}
       
       No. & Tool Name & Verification \& &  & Question Bank & Availability &	Relevant Study  \\
       
        &  & Abnormal Behavior & Security & Generation \& Evaluation	&  &  \\ \cline{1-7}
 
1 &	Secure Exam Environment (SEE) &	Yes & Yes &	Yes & N-A &	\cite{frankl2018pathways} \\ \cline{1-7}

2 & Unified E-Examination Solution & Yes & Yes & Yes & N-A & 	\cite{sabbah2017security} \\ \cline{1-7}

3 &	Examination Management System (EMS) & Yes & Yes & Yes & N-A & \cite{al2019integrated} \\ \cline{1-7}

4 & Continuous Online Authentication System & Yes & Yes & & N-A &	\cite{ghizlane2019new} \\ \cline{1-7}

5 &	Online Exam Proctoring (OEP) system & Yes &	Yes & & N-A & \cite{atoum2017automated} \\ \cline{1-7}

6 & Secure E-Learning System & Yes & Yes & & N-A & \cite{kausar2020fog} \\ \cline{1-7}

7 & Online Exam Authentication System &	Yes & - & -	& N-A & \cite{aisyah2018development} \\ \cline{1-7}

8 & Prototype Online Exam App & Yes & & & N-A & \cite{sukmandhani2019face} \\ \cline{1-7}

9 &	FLEXauth & Yes & & & N-A & \cite{opgen2018application} \\ \cline{1-7}

10 & Secure Online Examination System & & Yes & & N-A & \cite{mathapati2017secure} \\ \cline{1-7}

11 & Online Item Exam System & & & Yes & N-A & \cite{ketui2016item} \\ \cline{1-7}

12 & Exam Wizard & & & Yes & N-A & \cite{vomvyras2019exam} \\ \cline{1-7}

13 & Online Descriptive Answer Marking
System & & & Yes & N-A & \cite{das2019examination} \\ \cline{1-7}

14 & Clever Testing System & & & Yes & N-A & \cite{boussakuk2019online} \\ \cline{1-7}
15 & MoLearn System	& & & Yes & N-A	& \cite{lemantara2018prototype} \\ \cline{1-7}

16 & Snaptron & & & Yes & N-A & \cite{wagstaff2019automatic} \\ \cline{1-7}
17 & ViLLE & & & Yes & Public  & \cite{rajala2016automatically} \\ \cline{1-7}
18 & Simple and
Dynamic Examination System (SDES) & & & Yes & N-A & \cite{sultan2019automatically} \\ \cline{1-7}
19 & Automatic Evaluation System & & & Yes & N-A & \cite{ettarres2017evaluation} \\ \cline{1-7}

20 & e-Testing System & & & Yes & N-A & \cite{tashu2019intelligent} \\ \cline{1-7}
21 & Online Examination System & & & Yes & N-A & \cite{jiang2019design}
\\ \hline\hline
        \end{tabular}
\end{adjustbox}
    \vspace*{-5mm}
        \label{tab9:LeadingTools}
\end{table*}

Results of the proposed tools, as given in Table \ref{tab9:LeadingTools}, are really interesting. Overall, we identified 21 tools for online examination where 3 tools belong to Verification \& Abnormal Behavior feature, 1 tool belongs to Security feature and 11 tools belong to Question Bank Generation \& Evaluation feature. Moreover, 3 tools target both Verification \& Abnormal Behavior as well as Security features whereas 3 tools support all three online exams features. Therefore, it can be concluded that Question Bank Generation \& Evaluation is the frequently targeted feature in the proposed tools. It is important to note that the researchers claim the development of tool in few studies (e.g.~\cite{shi2017research} etc.). However, proper details about the proposed tools are missing in such studies. Therefore, we do not include such studies and their proposed tools in Table \ref{tab9:LeadingTools} due to lack of sufficient relevant information. For example, S. Prathish et al.~\cite{prathish2016intelligent} claim the development of intelligent system to monitor online exams where multi-modal biometrics are utilized. Authors explained the proposed approach properly, however, the details about the developed tool (e.g. interface, language / platform used for implementation etc.) are not provided. In addition to this, there are few studies particularly dealing with novel techniques / approaches without the development of tool. For example, Kassem et al.~\cite{kassem2017formal} proposed an interesting approach where formal methods are utilized to ensure the integrity of online exams. The development of tool with proper interface is usually not required for such proposals. Therefore, such types of studies are also not included in Table \ref{tab9:LeadingTools}. 

The results pertaining to the availability of proposed tools are really surprising. We found only one tool (i.e. ViLLE~\cite{rajala2016automatically}) where a web link\footnote{Accessible at: \url{https://ville.cs.utu.fi/old/?p=1}} is available to download few components (without actual source code). Other than ViLLE, rest of the tools proposed in studies under consideration do not provide any availability information (e.g. download link, source code etc.). Therefore, the proposed tools are of least significance for researchers and practitioners since further customization / extension or even evaluation is not possible. The tools availability results are really surprising and require further investigation. Therefore, we performed search (Google) for each proposed tool to find any additional information or web link. However, we were unable to find some proper download or source code link for any of the proposed tool. In fact, we only found very basic information about few proposed tools. For example, we found a login link~\cite{68examSystemLogin} for Online Item Exam System~\cite{ketui2016item}, where neither the language was known to us nor username and password was available due to which further evaluation of the system was also not possible.  Similarly, we found a web link~\cite{69OEP} where basic information regarding Online Exam Proctoring (OEP) system~\cite{atoum2017automated} was provided and relevant dataset was also available. However, it was not possible to download OEP or its code for evaluation purposes.

\subsubsection{Tools Utilized in Selected Studies}\label{utilizedtool}
So far, we presented the tools that have been proposed and developed as part of a research. However, it is equally important to highlight the existing tools that have been used in the selected studies for the implementation of proposed techniques and tools. This facilitates researchers and practitioner of the domain to select right tool as per requirements. Therefore, we present 25 important existing tools that have been used in the selected studies as given in Table \ref{tab10:ExistingTools}. The tool name is given in the second column and the purpose of tool is given in the third column of Table \ref{tab10:ExistingTools}. The relevant studies where the given tool is utilized are given in the last column.

Different programming languages have been used in the selected studies for the implementation of proposed technique / tool as given in Sr. \# 1 to 6 of Table \ref{tab10:ExistingTools}. The languages like Python and MATLAB are highly supported for the implementation of ML and AI techniques. Therefore, these languages are mostly utilized to implement ML / AI based approaches. For example, Sharma et al.~\cite{das2019examination} performed implementation of proposed AI based technique with Python where NLTK library is utilized for NLP operations. In another study, Atoum et al.~\cite{atoum2017automated} proposed ML based technique for cheating prevention where implementation (e.g. feature extraction, classification etc.) is carried out in MATLAB. In addition to implementation languages, different ML based tools and libraries like TensorFlow, OpenCV and Weka were also utilized in the selected studies as given in Table \ref{tab10:ExistingTools}. On the other hand, implementation languages like PHP, Java and C\# were mostly utilized in the selected studies for the development of a complete system / tool. For example, online exam assessment tool (Exam Wizard) is implemented in~\cite{vomvyras2019exam} with PHP. In another study~\cite{al2019integrated}, online exam management system is implemented in Java.

Several databases and storage platforms were utilized in the selected studies as given in Sr. \# 14 to 18 of Table \ref{tab10:ExistingTools}. MYSQL and Firebase cloud have been frequently utilized for storage purposes. In addition to storage platforms, there exist several special purpose tools that have been utilized in the selected studies to achieve particular objectives. For example, VirtualBox~\cite{andersen2020adapting} is a special tool to achieve virtualization that was utilized in two studies i.e.~\cite{natawiguna2016virtualization,frankl2018pathways}. Similarly, HiddenCameraActivity~\cite{chang2016review} provides image capturing features secretly and WeScan~\cite{kitchenham2004procedures} enables the effective scanning of documents. Furthermore, ProVerif~\cite{aisyah2018development} is a formal verification tool, which was used in~\cite{kassem2017formal} for the formal analysis of violations in online exams. To summarize, all aforementioned existing tools are utilized in the existing studies to achieve particular objective. It is important to mention that few studies did not provide any information about the languages and tools, which were used for implementation. For example, Mahatme et al.~\cite{mahatme2016data} proposed fuzzy logic-based approach for the intelligent classification of question bank, however, the information about the implementation language / tool was not given. Therefore, information about the utilized tools for such studies is not available in Table \ref{tab10:ExistingTools}. Similarly, in few studies (e.g.~\cite{hu2018research,diedenhofen2017pagefocus} etc.), very basic approaches like HTML, CSS, and JavaScript etc. were utilized for implementation and we therefore did not include this information in Table \ref{tab10:ExistingTools} for simplicity.

It is important to mention that Moodle is an open source E-learning platform~\cite{84moodle}, which is frequently utilized in different institutes throughout the world. However, Moodle does not support required online exams features like cheating prevention, which are necessary to ensure integrity of online exams. For this reason, few studies proposed online exam solutions, which are targeted to enhance the capabilities of Moodle for online exams. For example, Sabbah Y.W.~\cite{sabbah2017security} proposed a complete online exams framework and tool (i.e. ISEEU and SABBAH) that is implemented as a part of Moodle platform already deployed in the institute. Similarly, Yamna Ettarres~\cite{ettarres2017evaluation} proposed automatic evaluation technique and tool for online exams using Bayesian networks. Author integrated the proposed tool with Moodle, which was deployed in the University of Manouba, Tunisia. Such studies only perform the integration with Moodle, therefore, we did not include Moodle and corresponding studies in Table \ref{tab10:ExistingTools}.

\begin{table}[ht]
\caption{Existing Tools Utilized in the Selected Studies}
\vspace*{-3mm}
\renewcommand{\arraystretch}{1.5} 
\linespread{0.95}\selectfont
\centering
\begin{adjustbox}{max width=\columnwidth}
       \begin{tabular}{cllr}\\\hline
    
No. & Tool Name & Purpose & Relevant Studies \\\hline\hline
\multirow{2}{*}{1} &	\multirow{2}{*}{Python}  & \multirow{7}{*}{Implementation Languages}  &    \cite{asep2019design}, \cite{sukmandhani2019face}, \cite{garg2020convolutional},  \\
& & & \cite{das2019examination}, \cite{wagstaff2019automatic}, \cite{opgen2018application} \\ 

2 &	PHP     &   & \cite{ketui2016item}, \cite{vomvyras2019exam}, \cite{sabbah2017security}, \cite{ettarres2017evaluation}, \cite{tashu2019intelligent}  \\
3 &	Java    &   & \cite{shi2017research}, \cite{al2019integrated}        \\
4 &	Matlab  &   & \cite{atoum2017automated}, \cite{cote2016video} \\
5 &	C\#     &   & \cite{kausar2020fog} \\
6 &	C++     &   &  \cite{atoum2017automated} \\ \cline{1-4}

7 &	TensorFlow~\cite{70TensorFlow} &	Machine Learning Platform & \cite{asep2019design}, \cite{wagstaff2019automatic}	\\\cline{1-4}

8 &	Android Studio~\cite{71Android} & Android Dev. Platform & \cite{asep2019design}	\\\cline{1-4}

9 &	VirtualBox~\cite{72Vbox} & Virtualization & \cite{natawiguna2016virtualization}, \cite{frankl2018pathways} \\\cline{1-4}
10 &	Open CV~\cite{73openCV} & Image Processing and & \cite{asep2019design}, \cite{cote2016video}, \cite{garg2020convolutional}, \cite{traore2017ensuring} \\
11 &	Emgu CV~\cite{74emguCV}  &	Machine Learning Libraries & \cite{sukmandhani2019face} \\ \cline{1-4}

12 & NLTK~\cite{75nltk} & Natural Language Processing & \cite{das2019examination}\\
13 &	OpenNLP~\cite{76openNLP} & Libraries & \cite{kar2017novel} \\ \cline{1-4}

14 &	MySQL	    & \multirow{5}{*}{Databases / Storage}  & \cite{ketui2016item}, \cite{vomvyras2019exam}, \cite{sabbah2017security}, \cite{tashu2019intelligent}	\\
15 &	FireBase    &   & \cite{aisyah2018development}, \cite{vomvyras2019exam}, \cite{wagstaff2019automatic} \\
16 &	SQLlite	    &    & \cite{sukmandhani2019face}, \cite{das2019examination} \\
17 &	SQL Server  &   & \cite{shi2017research}, \cite{kausar2020fog}	\\
18 &	JSON        &   & \cite{aisyah2018development} \\ \cline{1-4}

19 &	Hidden Camera Activity~\cite{77hidden} & Image Capturing & \cite{aisyah2018development}\\ \cline{1-4}

20 &	Face++~\cite{78faceplus} & Facial Recognition Platform & \cite{shi2017research}\\ \cline{1-4}

21 &	WeScan~\cite{79weScan} &	Documents Scanning & \cite{wagstaff2019automatic}	\\ \cline{1-4}

\multirow{2}{*}{22} &	Bayesian Network tools & \multirow{2}{*}{Probability  Models Toolkit}	& \multirow{2}{*}{\cite{ettarres2017evaluation}} \\
&	in Java (BNJ)~\cite{80bayesian}  &	 &  \\\cline{1-4}

23 &	ProVerif~\cite{81prover} & Formal Verification Tool & \cite{kassem2017formal} \\ \cline{1-4}
24 &	Disco~\cite{82disco}  &	process mining toolkit & \cite{baykasouglu2018process} \\\cline{1-4}
25 &	Weka~\cite{83weka} &	Machine Learning Tool & \cite{chuang2017detecting} \\

\hline\hline
        \end{tabular}
\end{adjustbox}
    \vspace*{-5mm}
        \label{tab10:ExistingTools}
\end{table}

\subsection{Datasets}\label{data}
Datasets are really important for reliable validation of a proposed technique / tool. Therefore, trustworthy datasets are essential while authenticating the outcomes of a proposal. We identified 11 datasets used / proposed in the selected studies for validation as given in Table \ref{tab11:leadingDatasets}. The dataset name is given in second column and characteristics of dataset are given in third column of Table \ref{tab11:leadingDatasets}. Characteristics of dataset include format (Video, audio, text etc.), number of records and purpose (i.e. targeted online exam feature through dataset). The availability of dataset (i.e. Public, Private and Not-Applicable – N-A) is given in fourth column of Table \ref{tab11:leadingDatasets}. Finally, the reference of relevant study where the given dataset is actually utilized / proposed, is given in last column.

We found six publicly available datasets as given in Sr. \# 1 to 6 of Table \ref{tab11:leadingDatasets}. Out of these six public datasets, only Online Exam Proctoring (OEP) dataset was newly constructed in~\cite{atoum2017automated} whereas the rest were benchmark datasets that were reused by~\cite{hu2018research,wagstaff2019automatic,traore2017ensuring,kausar2020fog,wu2020exam}. The reference of each publicly available dataset is provided against the name (second column) for further investigation. On the other hand, we identified five datasets (Sr. \# 7 to 11 of Table \ref{tab11:leadingDatasets}) where the availability information (e.g. download link etc.) was missing and therefore represented as N-A (Not Applicable) in the table. For example, Ketui et al.~\cite{ketui2016item} developed a dataset for validation by utilizing different existing online exam items like teacher assistance, government, and company exams. However, the details of developed dataset were not properly explained and availability information (e.g. download link etc.) was totally missing. Likewise, authors in~\cite{cote2016video} developed a dataset comprising of 6 videos with 25311 frames, however, the availability details were totally missing. In another study~\cite{chen2018application}, authors claimed the development of a dataset, but relevant details including total number of records were missing.

It can be analyzed from Table \ref{tab11:leadingDatasets} that six datasets have been utilized for Verification \& Abnormal Behavior feature, four datasets for Question Bank Generation \& Evaluation feature and only one dataset has been employed for Security feature. Among these, five datasets are based on textual format whereas three datasets comprise images. Furthermore, two datasets are based on video format and only one dataset contains both audio and video as given in Table \ref{tab11:leadingDatasets}. It is important to mention that other selected studies performed different types of experiments, surveys, and test scenarios for the validation of proposal without employing particular dataset. For example, Teemu Rajala et al.~\cite{rajala2016automatically} validated the proposed approach through the participation of 478 students with four exam occurrences. In another study~\cite{tashu2019intelligent}, survey comprising 20 teachers has been conducted to validate the usability of the proposed system.

\begin{table*}
\caption{Leading Datasets Used / Proposed in the Selected Studies}
\vspace*{-3mm}
\renewcommand{\arraystretch}{1.5} 
\linespread{0.95}\selectfont
\centering
\begin{adjustbox}{max width=\textwidth}
       \begin{tabular}{cllllcr}\\\hline
    
      & &\multicolumn{3}{c}{Characteristics} & & Relevant \\\cline{3-5}
       
      No. & Dataset Name & Format &  No of Records & Purpose & Availability &	Study  \\ \hline\hline

1 &	Online Exam Proctoring (OEP) &	Audio and Video	& 72 Files &	Verification \& Abnormal Behavior &	Public & \cite{atoum2017automated} \\\cline{1-7}

2 &	AFLW~\cite{85facial} &	Images & 21997 & Verification \& Abnormal Behavior & Public & \cite{hu2018research} \\ \cline{1-7}
3 &	EMNIST~\cite{86emnist} & Images & 6 datasets group	& Question Bank Generation \& Evaluation & Public & \cite{rajala2016automatically} \\\cline{1-7}

4 &	Extended Yale Face Database~\cite{87yale} & Images & 16128 & Verification \& Abnormal Behavior & Public & \cite{traore2017ensuring} \\\cline{1-7}

5 &	OULAD~\cite{88analytics} & Text & 7 CSV files & Security & Public & \cite{kausar2020fog} \\\cline{1-7}

6 &	ASSISTments~\cite{89assist} & Text & 3 CSV files & Question Bank Generation \& Evaluation & Public & \cite{wu2020exam} \\\cline{1-7}
7 &	Online Exams Items & Text & 1000 items & Question Bank Generation \& Evaluation & N-A & \cite{ketui2016item} \\\cline{1-7}
8 &	Video Dataset & Video & 6 Videos (25311 Frames) & Verification \& Abnormal Behavior & N-A	& \cite{cote2016video}\\\cline{1-7}
9 &	Video Dataset &	Video &	30 Videos &	Verification \& Abnormal Behavior & N-A & \cite{prathish2016intelligent} \\\cline{1-7}
10 & Questions Dataset & Text & N-A	& Question Bank Generation \& Evaluation & N-A	& \cite{chen2018application} \\\cline{1-7}
11 & Student Dataset & Text & 156 docs & Verification \& Abnormal Behavior & N-A & \cite{opgen2018application} \\\cline{1-7}
\hline\hline
        \end{tabular}
\end{adjustbox}
    \vspace*{-5mm}
        \label{tab11:leadingDatasets}
\end{table*}

\subsection{Country of Research}\label{researchCountry}
To analyze the participation of countries in online exams research, the selected studies were thoroughly investigated to identify the contributing institutes and corresponding countries. We identified 25 countries that contributed to online exams research as given in Table \ref{tab12:ResearchCountry}. It is analyzed that china and India are the leading contributors in online exams research as 8 studies belong to each of them. Furthermore, Indonesia is the third leading contributor on the list with 7 studies where Teknologi Bandung, Indonesia is a leading institute with 3 studies.

\begin{table}[ht]
\caption{Country and Institution participated in the selected studies}
\vspace*{-2mm}
\renewcommand{\arraystretch}{1.15} 
\linespread{0.95}\selectfont
\centering
\begin{adjustbox}{max width=\columnwidth}
       \begin{tabular}{cllr}\\\hline

No. & Country & University / Institute & Relevant Studies  \\\hline\hline

1 &	Austria & 	Alpen-Adria-Universität Klagenfurt &	\cite{frankl2018pathways} \\ \cline{1-4}
2 &	Canada &	University of Victoria, Victoria, BC &	\cite{cote2016video},\cite{traore2017ensuring} \\ \cline{1-4}

\multirow{8}{*}{3} &	\multirow{8}{*}{China} & College of Infor. Science and Techn. Dalian, Liaoning &	\cite{hu2018research} \\ \cline{3-4}
	& &	Huaihai Institute of Technology Lianyungang Jiangsu & \cite{shi2017research}\\ \cline{3-4}
	& &	Huazhong University of Science and Technology &	\cite{fan2016gesture}\\ \cline{3-4}
	& &	South China University of Technology, Guangzhou &	\cite{chen2018application}\\ \cline{3-4}
	& &	South China Normal University, Guangzhou &	\cite{wu2020exam}\\ \cline{3-4}
	& &	Bijing Institute of Technology & \cite{zhang2018analysis}\\ \cline{3-4}
	& &	Guangdong University of Foreign Studies	& \cite{jiang2019design}\\ \cline{3-4}
	& &	Shanghai University &	\cite{kausar2020fog} \\ \cline{1-4}

4 &	Finland &	University of Turku Graduate School (UTUGS) &	\cite{rajala2016automatically}\\ \cline{1-4}
5 &	France & Université Grenoble Alpes, INRIA &	\cite{kassem2017formal}\\ \cline{1-4}
6 &	\multirow{2}{*}{Germany} & RWTH Aachen University &	\cite{opgen2018application} \\ \cline{3-4}
		& & University of Düsseldorf	& \cite{diedenhofen2017pagefocus}\\ \cline{1-4}
		
7 &	Greece &	Department of Informatics University of Piraeus &	\cite{vomvyras2019exam}\\ \cline{1-4}
8 &	Hungary & ELTE-Eötvös Loránd University, Budapest &	\cite{tashu2019intelligent} \\ \cline{1-4}
\multirow{5}{*}{9} & \multirow{5}{*}{Indonesia} & Institute Teknologi Bandung &	\cite{asep2019design},\cite{aisyah2018development},\cite{natawiguna2016virtualization} \\ \cline{3-4}
	& &	Bina Nusantara University &	\cite{sukmandhani2019face} \\ \cline{3-4}
	& &	Udayana University Denpasar, Bali &	\cite{sukadarmika2018introducing} \\ \cline{3-4}
	& &	Institute of Business and Informatics Stikom Surabaya &	\cite{lemantara2018prototype} \\ \cline{3-4}
	& &	Universitas Brawijaya,  Malang & \cite{fanani2019interactive} \\ \cline{1-4}
\multirow{7}{*}{10} & \multirow{7}{*}{India}	& Amrita School of Engineering &	\cite{prathish2016intelligent},\cite{subramanian2018using} \\ \cline{3-4} & &	Medi-Caps University (MU) &	\cite{garg2020convolutional} \\ \cline{3-4}
	& &	ACS College of Engineering, Bangalore &	\cite{mathapati2017secure} \\ \cline{3-4}
	& &	Kavikulguru Institute of Technology \& Science &	\cite{mahatme2016data} \\ \cline{3-4}
	& &	Anna University, Chennai &	\cite{nandini2020automatic} \\ \cline{3-4}
	& &	University Bhubaneswar &	\cite{das2019examination} \\ \cline{3-4}
	& &	NITTTR Kolkata &	\cite{kar2017novel} \\ \cline{1-4}
11 &	Jordon &	German Jordanian University, Amman & \cite{al2019integrated} \\ \cline{1-4}
12 &	Kuwait &	Kuwait University &	\cite{sultan2019automatically} \\ \cline{1-4}
13 &	Malaysia & 	Universiti Kebangsaan Malaysia	& \cite{karim2016proposed} \\ \cline{1-4}
\multirow{2}{*}{14} & \multirow{2}{*}{Morocco} &	ESTC, Hassan II University, Casablanca & \cite{ghizlane2019new} \\ \cline{3-4} 
	& &	Sidi Mohamed Ben Abdellah University Fez & \cite{boussakuk2019online} \\ \cline{1-4}
15 &	New Zealand & University of Auckland & \cite{manoharan2019cheat} \\ \cline{1-4}
16 &	Palestine &	Al-Quds Open University, Ramallah &	\cite{sabbah2017security} \\ \cline{1-4}
\multirow{2}{*}{17} & \multirow{2}{*}{Philippines} & Lyceum of the Philippines University & \cite{chua2019online} \\ \cline{3-4}
		& & Techn. Institute of the Philippines Cubao, Quezon City & \cite{abisado2018towards} \\ \cline{1-4}
18 &	Romania &	University of Craiova &	\cite{albastroiu2018exam} \\ \cline{1-4}
19 &	Russia 	& ITMO University, Saint-Petersburg & \cite{matveev2020virtual} \\ \cline{1-4}
20 &	Saudi Arabia & 	Prince Sattam Bin Abdulaziz University &	\cite{kolhar2018online} \\ \cline{1-4}
21 &	Thailand &	Rajamangala University of Technology Lanna, Nan & \cite{ketui2016item} \\ \cline{1-4}
22 &	Tunisia &	University of Manouba &  \cite{ettarres2017evaluation} \\ \cline{1-4}
23 &	Turkey 	& Dokuz Eylul University, Izmir &  \cite{baykasouglu2018process} \\ \cline{1-4} 
24 &	UK &	Cardiff Metropolitan University &	\cite{ullah2019dynamic} \\ \cline{1-4}
\multirow{5}{*}{25} & \multirow{5}{*}{USA} &	Michigan State University &	\cite{atoum2017automated} \\ \cline{3-4}
	& & School of Business, Hampton University & \cite{d2017conceptual} \\ \cline{3-4}
	& & Arizona State University &	 \cite{chuang2017detecting} \\ \cline{3-4}
	& &	UCLA, Los Angeles &	 \cite{wagstaff2019automatic} \\ \cline{3-4}
	& &	University of Memphis &	\cite{golden2020addressing} \\
\hline\hline
        \end{tabular}
\end{adjustbox}
    \vspace*{-5mm}
        \label{tab12:ResearchCountry}
\end{table}

\begin{figure}[hbt!]
    \centering
    \includegraphics[]{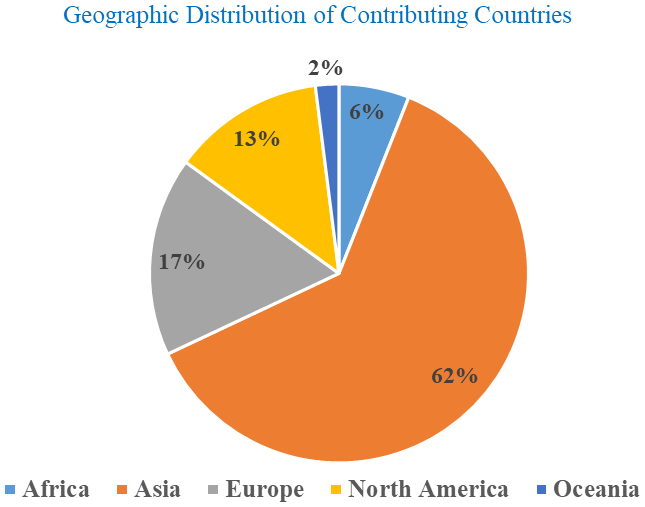}
    \caption{Summary of online exams participation on the basis of five major continents}
    \label{fig4:ContinentParticipation}
\end{figure}

The summary of geographical distribution of online exams contributions, on the basis of five major continents, is shown in Fig. \ref{fig4:ContinentParticipation}. It can be analyzed that Asia is the leading contributor with 62\% (33 studies) of overall conducted research. The closest competitor is Europe that stands with 17\% (9 studies) on the list. Similarly, North America and Africa stand fourth and fifth on the list with 13\% (7 studies) and 6\% (3 studies), respectively. Oceania (New Zealand) is last standing on the list with merely 2\% of the selected publications. In short, most of the online exams research has been conducted in Asian countries during past five years.

\section{Key Factors for Global Adoption}\label{keyFactors}
\begin{figure}[hbt!]
    \centering
    \includegraphics[]{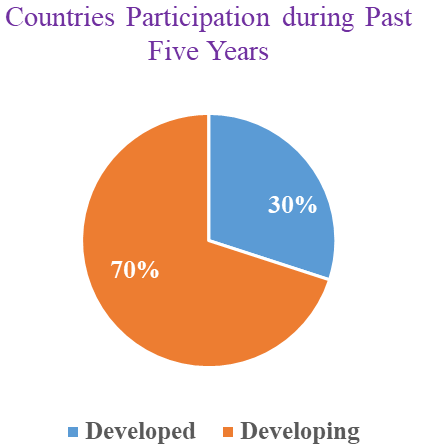}
    \caption{Participation of online exams research with respect to developed and developing countries}
    \label{fig5:CountriesParticipation}
\end{figure}

To this point, we have analyzed different features, underlying development approaches and techniques / algorithms in the domain of online exams. Further, various tools proposed by a number of researchers for online exams have also been explored. Therefore, on the basis of our analysis, we now try to identify and investigate important factors for the adoption of online exams globally. In this context, the main countries participated in online exams research are identified in Section \ref{researchCountry}. On the basis of Table \ref{tab12:ResearchCountry} highlights (Section \ref{researchCountry}), the participating countries can be classified into two groups i.e. Developed and Developing Countries as per International Monetary Fund (IMF) organization~\cite{90monetary}. Particularly, developed countries have excellent financial and economic status whereas developing countries have struggling economies with low financial values. Consequently, developed countries have more funds and stable infrastructure to support online exams as compared to developing countries. The overall participation in the selected studies (Table \ref{tab12:ResearchCountry}) on the basis of developed and developing countries is shown in Fig. \ref{fig5:CountriesParticipation}.

It can be analyzed from Fig. \ref{fig5:CountriesParticipation} that the 70\% (37 studies) of the contributions in online exams are from developing countries during past five years. On the other hand, 30\% (16 studies i.e.~\cite{atoum2017automated,cote2016video,vomvyras2019exam,wagstaff2019automatic,rajala2016automatically,opgen2018application,traore2017ensuring,frankl2018pathways,tashu2019intelligent,kassem2017formal,ullah2019dynamic,diedenhofen2017pagefocus,golden2020addressing,manoharan2019cheat,d2017conceptual,chuang2017detecting}) contributions  came from developed countries. Therefore, it seems that the adoption of online exams is more frequent in developing countries as compared to developed countries. However, this is not the actual case. On the basis of detailed analysis of selected studies, it is examined that the contributions came from developing countries are mostly theoretical without actual practicability. Therefore, most of the research from developing countries is only good for educational purposes and cannot be actually applied in real online exams environment. For example, the studies like~\cite{das2019examination,subramanian2018using} from India only proposed theoretical frameworks, which are difficult to implement in real environment especially in existing unstable educational infrastructure of the country. On the other hand, most of the contributions from developed countries are practical and can be applied in real online exams environment due to the stable educational infrastructure. For example, the study~\cite{atoum2017automated} from Michigan State University (MSU) USA proposed a complete and automated proctoring solution for cheating prevention in online exams. As a part of research, a dataset is developed, which is now publicly available for further investigation. The solution is based on several biometrics with real time monitoring and its viability is established through real online exams environment. The results proved that the proposed solution is highly effective for cheating prevention in online exams and can be applied in real environment. To summarize, the proposals like~\cite{atoum2017automated} are actually implemented in developed countries. On the other hand, such proposals (e.g.~\cite{atoum2017automated} cannot be applied in developing countries due to financial issues and unstable infrastructure. For example, the solution in [16] performs real time monitoring of examinee’s face movements, which requires high internet speed and stable network infrastructure. However, stable network platform with good internet speed is usually not available in most of the developing countries. It is important to mention that the countries like China are in the list of developing countries though, they have advanced E-learning platform and better financial resources. Therefore, the contributions of such countries are also more practical as compared to other least developing countries. 

On the basis of aforementioned analysis, it can be concluded that the financial situations and existing E-learning infrastructure are highly important for the adoption of online exams. For this reason, a single or a set of particular approaches cannot be applied for online exams globally. Therefore, there is a need for the investigation of important factors with respect to online exams attributes, so that different countries can adopt online exams with appropriate features as per financial and existing infrastructure support. In this regard, we have identified four important online exams adoption factors as follows:

\begin{enumerate}
    \item \textbf{Network Infrastructure:} This factor refers to the overall network infrastructure of a particular country where availability, consistency and speed are important characteristics. Particularly, Infrastructure is considered as “Excellent” if internet is available to all examinees and invigilators on different geographical locations in a particular country. The speed of internet is high (i.e. between 50 MB to 100 MB per second) and consistent. The Infrastructure is considered as “Good” if internet is available to all examinees and invigilators, however, internet speed varies between 10 MB to 20 MB per second and continuous connectivity is consistent. The Infrastructure is considered as “Low” if internet is not available to all examinees and invigilators. Furthermore, internet speed varies between 2MB to 10 MB per second and continuous connectivity is inconsistent. 
    \item \textbf{Hardware Requirements:} This factor refers to the number of hardware requirements, based on financial conditions, supported by a particular country. The hardware may include computers, servers, cameras etc. for the seamless execution of online exams. The hardware requirements can be considered as “Large” if higher costs are required for procurement. Moreover, hardware requirements can be considered as “Average” in case the procurement can be managed in reasonable cost. Finally, the hardware requirements can be “Small” if managed in a lower cost.  
    \item \textbf{Implementation Complexity:} This factor refers to the involvement of complexity for the development of a particular online exam solution. Implementation complexity is directly linked with the cost of online exam i.e. higher implementation complexity leads to higher cost. The implementation complexity can be referred as High, Medium, and Low. Particularly, a complete online exam solution, developed through machine learning / artificial intelligence techniques, usually leads to high or medium implementation complexity. On the other hand, implementation complexity is low for a typical online exam solution, which is developed through traditional languages like Java and PHP.
    \item \textbf{Training Requirements:} This factor refers to the involvement of any training requirements, which are required for the execution of online exams. Training requirements may belong to examinee and / or invigilator. For example, the students with computer science / information technology background are familiar with the usage of online exam systems. However, the students with other subjects like political science, accounting etc. may not be able to operate complex online exam system. For such students, it may be required to perform training before actual exam. On the other hand, invigilator may also require training in case of complex ML/AI based system dealing with automatic assessment of online exams. Training requirements directly affect the online examination cost. Training requirements can be classified as High and Low.

\end{enumerate}

Certain characteristics of the aforementioned factors are highly important while adopting particular online exams features in real environment. These factors are directly linked with the system’s overall cost, which is a major concerning element for most of the developing countries. In this context, it is required to investigate the requirements and effects of leading online exams features (Section \ref{leadingAttribute}) on each factor. This facilitates the selection of a right online exam system for a particular country on the basis of existing E-learning infrastructure and overall cost. Therefore, we performed comparative analysis of key online exam features with respect to adoption factors as given in Table \ref{tab13:KeyAdoptionFactors}. The important online exam features are given in first column of Table \ref{tab13:KeyAdoptionFactors}. Moreover, four important factors (i.e. Network Infrastructure, Hardware Requirements, Implementation Complexity and Training Requirements) along with aforementioned characteristics are given in second, third, fourth and fifth columns of Table \ref{tab13:KeyAdoptionFactors}, respectively. Finally, respective overall cost is given in last column.

To evaluate the requirement / effect of each feature with respect to a particular factor, three symbols / abbreviations are utilized. The tick symbol (\checkmark) represents that a specified characteristic of a given factor is sufficient for the implementation of a particular online exam feature. On the other hand, the cross symbol (×) represents that  a given online exam feature cannot be implemented through the specified characteristic of a factor. Finally, the essential characteristic of a particular factor, which is at least required for the implementation of a given feature is represented through Mandatory - (M) abbreviation as shown in Table \ref{tab13:KeyAdoptionFactors}.

It is important to note that four key online exams attributes are already defined in Section \ref{leadingAttribute}. Here, each attribute is logically divided into two groups in order to perform realistic comparative analysis as shown in Table \ref{tab13:KeyAdoptionFactors}. For example, Verification \& Abnormal Behavior attribute is divided into Biometric Based and Application Based groups where Biometric approaches (e.g.~\cite{atoum2017automated}) utilize examinee images, videos etc. to evaluate examinee verification and / or abnormal behavior. On the other hand, the application-based approaches (e.g.~\cite{diedenhofen2017pagefocus}) utilize examinee system’s events (e.g. Browser Window Status etc.) to detect cheating abnormalities. In addition to this, Security attribute is classified as basic and advanced where advanced security provides complete and secure communication mechanism between all servers and clients of online exam system (e.g.~\cite{kausar2020fog}). In the same way, Question Bank Generation \& Evaluation attribute is divided into two groups (i.e. ML / AI based and Traditional) where the ML / AI based approaches (e.g.~\cite{chen2018application}) apply advanced techniques for the generation and assessment of question bank. In contrast, Traditional approaches (e.g.~\cite{vomvyras2019exam}) utilize different languages (e.g. Java, PHP etc.) without employing any latest ML / AI techniques. Finally, Usability attribute can be divided as Good or Fair where Good usability (e.g.~\cite{karim2016proposed}) significantly improves the interaction of examinee with a given online exam system.

\begin{table*}
\caption{Comparative analysis of online exams features with respect to key adoption factors}
\vspace*{-3mm}
\renewcommand{\arraystretch}{1.5} 
\linespread{0.95}\selectfont
\centering
\begin{adjustbox}{max width=\textwidth}
       \begin{tabular}{lcccccccccccc}\\\hline
    
\multirow{2}{*}{Online Exam Feature} & \multicolumn{3}{c}{Network Infrastructure} & \multicolumn{3}{c}{Hardware Requirements} & \multicolumn{3}{c}{Implementation Complexity} & \multicolumn{2}{c}{Training Requirements} & \multirow{2}{*}{Overall Cost}\\\cline{2-12}
       
 & Excellent & Good & Low & Large & Average & Small & High & Medium & Low & High & Low &   \\ \hline\hline

\multicolumn{13}{c}{\textbf{Verification \& Abnormal Behavior}}  \\ \cline{1-13}

Biometric Based	& \checkmark & M &  × &	\checkmark &	M &	× & \checkmark &	M &	× &	M &	× &	High \\ \cline{1-13}
Application Based &	\checkmark & \checkmark &	M &	\checkmark &	\checkmark &	M &	\checkmark &	\checkmark &	M & \checkmark &	M	& Low \\ \cline{1-13}

\multicolumn{13}{c}{\textbf{Security}}  \\ \cline{1-13}

Basic &	\checkmark &	\checkmark &	M &	\checkmark	& \checkmark &	M &	\checkmark & \checkmark &	M &	\checkmark &	\checkmark	& Low \\ \cline{1-13}
Advanced	& \checkmark &	M &	× &	\checkmark &	M &	× &	\checkmark &	M &	× &	\checkmark &	M &	High \\ \cline{1-13}

\multicolumn{13}{c}{\textbf{Question Bank Generation \&  Evaluation}}  \\ \cline{1-13}

ML / AI Based &	\checkmark &	M &	× &	\checkmark &	\checkmark &	M &	\checkmark &	M &	× &	M &	\checkmark &	High \\ \cline{1-13}
Traditional &	\checkmark &	\checkmark &	M &	\checkmark &	\checkmark &	M &	\checkmark &	\checkmark &	M &	\checkmark &	M &	Low \\ \cline{1-13}
 
\multicolumn{13}{c}{\textbf{Usability}}  \\ \cline{1-13}
Good &	\checkmark &	 \checkmark &	M &	\checkmark &	\checkmark &	\checkmark &	\checkmark &	\checkmark &	M &	\checkmark &	\checkmark &	Low \\ \cline{1-13}
Fair &	\checkmark & \checkmark &	M &	\checkmark &	\checkmark &	\checkmark &	\checkmark &	\checkmark &	\checkmark	& \checkmark &	M &	Low \\ 
\hline\hline
        \end{tabular}
\end{adjustbox}
    \vspace*{-5mm}
        \label{tab13:KeyAdoptionFactors}
\end{table*}

It can be analyzed from Table \ref{tab13:KeyAdoptionFactors} that biometric based Verification \& Abnormal Behavior feature of online exams at least requires good Network Infrastructure. Particularly, biometric based solutions (e.g.~\cite{atoum2017automated}) require real time monitoring of huge data (e.g. videos, images, voice etc.), which cannot be accomplished through low network infrastructure. Therefore, at least good network infrastructure is essential for such solutions. In addition to this, different types of hardware components (e.g. Cameras, streaming / processing severs etc.) are required for the implementation of biometric solutions. Therefore, average hardware requirements are at least compulsory and biometric solutions cannot be implemented through low hardware requirements. In addition, the biometric solutions usually require the application of ML / AI approaches for real time verification and detection of abnormalities. Therefore, medium level implementation complexity is at least involved for such solutions. As biometric solutions deal with different types of analyses and monitoring tasks, the invigilators usually require  intensive training for the proper execution of such solutions. Therefore,  training requirements for biometric solutions are usually high. On the basis of overall comparative analysis, it can be concluded that higher costs are required for the implementation of biometric solutions as given in Table \ref{tab13:KeyAdoptionFactors}.

On the other hand, Verification \& Abnormal Behavior feature can be achieved through Application based approaches with low network infrastructure. Particularly, application-based solutions do not require the transfer of huge real time data between servers and clients. A typical example of such solution is PageFocus~\cite{diedenhofen2017pagefocus} where the behavior of examinee’s browser window is analyzed to detect cheating abnormalities. Similarly, application-based solutions do not require large hardware requirements and system can be operational with minimum hardware. Furthermore, traditional languages like PHP, JavaScript etc. can be utilized for system development without employing any special ML / AI approaches. Therefore, application-based solutions can be implemented with low implementation complexity as given in Table \ref{tab13:KeyAdoptionFactors}. In addition to this, application-based solutions usually do not entail intensive training requirements and system can be operational with basic training. Finally, the overall cost of application-based systems is much lower than the biometric based solution.

The basic security feature in online exams can be achieved with low network infrastructure and small hardware requirements. It can be attained with low implementation complexity without employing any special trainings. On the other hand, advanced security aspects of online exams usually require  good network infrastructure as continuous internet availability is important. Moreover, different types of firewalls, servers may be required for highly secured systems, therefore, advanced security may have medium hardware requirements. Furthermore, medium level implementation complexity is required for advanced security and basic training of network / system engineer may also be needed for the execution of system. On the basis of security feature analysis, it can be concluded that the implementation of basic and advanced security in online exams may require lower and higher overall costs, respectively.

The Question Bank Generation \& Evaluation is highly important feature in online exams. This feature can be attained through two approaches i.e. ML / AI based and Traditional. Particularly, ML / AI based techniques perform  real time generation and assessment of question bank intelligently during online exams. Therefore, good network infrastructure is mandatory for ML / AI based approaches. However, such approaches do not usually require any special hardware and can be operational with limited hardware requirements. In addition, these approaches employ advanced ML / AI techniques, therefore, medium level implementation complexity is usually involved, and thus intensive system training is also required. Overall, ML / AI based Question Bank Generation \& Evaluation approaches require higher costs. On the other hand, traditional approaches utilize common system development languages like Java, PHP etc. without employing any ML / AI methodology. Further, real time generation and assessment of questions is not usually performed. Consequently, traditional approaches can operate in low network infrastructure with minimum hardware requirements. Moreover, the implementation complexity of such approaches is also low. These approaches can operate without performing any major training. Finally, the overall cost of such approaches is relatively low as compared to ML / AI based approaches.

Usability feature in online exam systems typically is not directly linked with network infrastructure and hardware requirements. However, to ensure the availability of online exam system, usability feature may require low network infrastructure. On the other hand, any particular hardware requirements are not usually involved, and good usability can be achieved even with very limited hardware. Particularly, Good usability can be achieved with low implementation complexity while fair usability is naturally evolved during system development without employing any special implementation strategy. In addition to this, Good usability leads to simple and self-explanatory execution of online exam system where examinee and invigilator training is not usually required. On the other hand, basic training of a system is commonly required in case of fair usability. Overall, the lower costs are involved while achieving good usability in online exams systems.

The analysis of major online exam features with respect to key factors, as given in Table \ref{tab13:KeyAdoptionFactors}, is a significant step towards the adoption of online exams globally. Given the facts in Table \ref{tab13:KeyAdoptionFactors}, the institutes and countries may initiate the online exams systems on the basis of their existing E-learning infrastructure and availability of funds. For example, the least developing countries (e.g. Bhutan, Yemen etc.) with significantly low income can startup online exams systems with following features: 
\begin{enumerate}
    \item Verification \& Abnormal Behavior feature through Application based approach
    \item Basic Security 
    \item Question Bank Generation \& Evaluation feature through traditional approach
    \item Fair usability.
\end{enumerate}
Particularly, these online exam features require basic network infrastructure and hardware requirements. Therefore, online exam system with such features can be deployed with minimal cost in least developing countries. Although such a system may not provide required online exams integrity, it is still a good starting point for least developing countries. Later on, further improvements can be made in the system on the basis of available resources and funds. Similarly, other developing and developed countries can deploy a proper online exam system with appropriate features, on the basis of available infrastructure, resources, and funds, by utilizing the facts of Table \ref{tab13:KeyAdoptionFactors}.            

It is important to note that analysis performed in Table \ref{tab13:KeyAdoptionFactors} is based on general observations, which are derived from the investigation of selected studies. For example, the observation like “biometric based studies usually require good network infrastructure” is based on the facts, which are given in several selected studies e.g.~\cite{atoum2017automated,cote2016video} etc. Similar is the case with other observations. Therefore, the analysis performed in Table \ref{tab13:KeyAdoptionFactors} is authentic. In fact, it is a significant step towards the global adoption of online exams.    
\section{Answers to RQs and Limitations}
To this point, the selected studies are thoroughly investigated and required results are precisely presented in Section \ref{result}. Furthermore, the key factors for the global adoption of online exams are identified and comparative analysis is also performed in Section \ref{keyFactors}. Consequently, we are now able to provide authentic answers to the research questions as follows:
\begin{figure}[hbt!]
    \centering
    \includegraphics[]{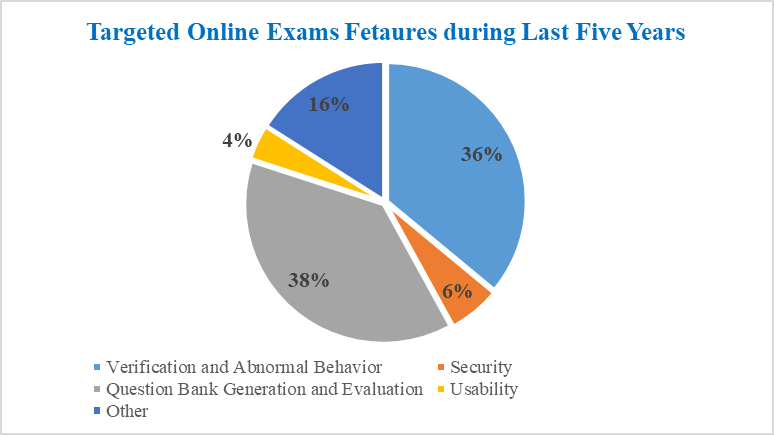}
    \caption{Summary of targeted Features in Selected Studies}
    \label{fig6:TargettedFeatures}
\end{figure}

\begin{figure}[hbt!]
    \centering
    \includegraphics[]{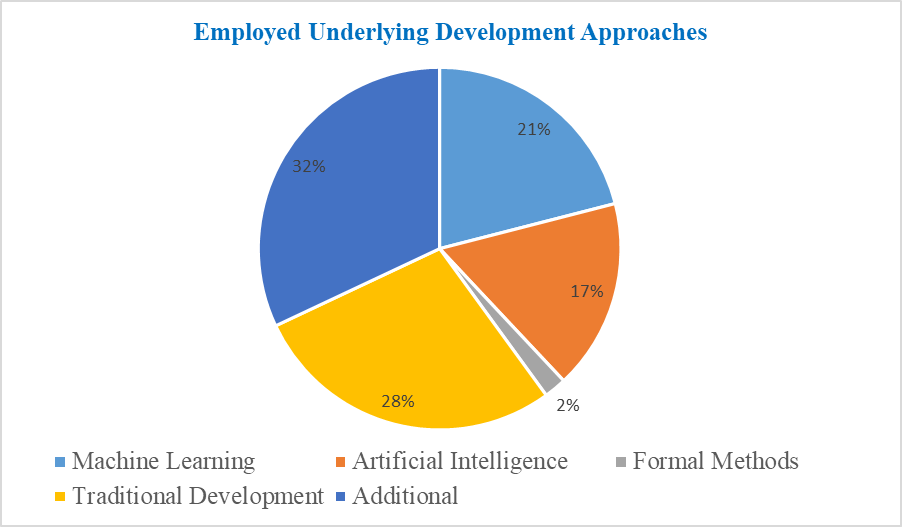}
    \caption{Summary of underlying development approaches utilized for online exams solutions during last five years}
    \label{fig7:LastDevelopmentApproaches}
\end{figure}

\begin{questions}
        \item What are the leading studies particularly dealing with the online exams solutions during past five years i.e. 2016-2020?
        
        \noindent \textbf{Answer:} Overall, 53 studies dealing with different aspects of online exams have been identified from six renowned databases i.e.  IEEE, Elsevier, Springer, ACM, Wiley, and Taylor \& Francis. The distribution of selected studies with respect to databases is provided in Table \ref{tab3:SummaryPublisher}. The distribution of studies with respect to publication year is provided in Table \ref{tab2:yearWise}. The distribution of studies with respect to publication type (i.e. conference or Journal) is given in Fig. \ref{fig3:StudiesDistribution}. The classification of selected studies with respect to major categorizes (i.e. Biometric, Software Applications and General) is performed in Section \ref{category} (Table \ref{tab5:primarycategory}).
        \item What are the major online exams features reported in the literature during past five years? 
        
        \noindent \textbf{Answer:} The four leading online exams attributes (i.e. Verification \& Abnormal Behavior, Security, Question Bank Generation \& Evaluation and Usability) are defined in Section \ref{leadingAttribute}. Subsequently, the grouping of selected studies on the basis of leading features is performed in Table \ref{tab6:leadingExamFeatures}. The summary of targeted features, based on the results of Section \ref{leadingAttribute}, is shown in Fig. \ref{fig6:TargettedFeatures}. It is concluded from Fig. \ref{fig6:TargettedFeatures} that Question Bank Generation \& Evaluation and Verification \& Abnormal Behavior are the most frequently targeted features with 38\% (20 studies) and 36\% (19 studies), respectively. Moreover, the frequency of Security and Usability features is 6\% (3 studies) and 4\% (2 studies), respectively. Furthermore, 16\% studies (9) are also identified in other class where more than one aforementioned feature are targeted simultaneously.
        
        \item What are the main underlying development approaches that have been employed for the implementation of online exams solutions? 
        
        \noindent \textbf{Answer:} The five main classes pertaining to underlying development approaches are defined in Section \ref{developmentApproach}. Subsequently, results are presented in Table \ref{tab7:developmentApproaches}. The summary of results, based on Table \ref{tab7:developmentApproaches}, is given in Fig. \ref{fig7:LastDevelopmentApproaches}. It can be concluded that traditional development languages like Java, PHP, C\# etc. are most commonly utilized for online exams solutions as 28\% researches (15 studies) belong to it. The ML and AI techniques are utilized in 21\% (11 studies) and 17\% (9 studies) researches, respectively. Furthermore, only one study (2\%) utilized formal methods in the area of online exams. It is important to note that the Additional group includes studies where both traditional development and ML / AI techniques have been combined to propose some sophisticated online exam solution. Furthermore, Additional group also incorporates studies, which cannot be fit in other groups e.g. theoretical framework etc. Overall, 32\% (17 studies) researches belong to Additional group as shown in Fig. \ref{fig7:LastDevelopmentApproaches}.

        \item What are the leading techniques / algorithms proposed in the domain of online exams?
        
        \noindent \textbf{Answer:} We identified 16 leading techniques / algorithms, which have been proposed by researchers in the selected studies as given in Section \ref{techniqueAlgo} (Table \ref{tab8:LeadingTechniques}). It is analyzed from the results that most of the proposed techniques and algorithms are based on the concepts of Machine Learning and Artificial Intelligence. Further details are available in Section \ref{techniqueAlgo}. 

        \item What are the major online exams tools proposed in the literature and how existing tools are utilized in the online exams research?
        
        \noindent \textbf{Answer:} We overall identified 21 tools that have been proposed by researchers in the selected studies as given in Section \ref{proposedtool} (Table \ref{tab9:LeadingTools}). Particularly, three proposed tools provide all major online exams features (i.e. Verification \& Abnormal Behavior, Security, Question Bank Generation \& Evaluation) simultaneously. Moreover, another three tools provide both Verification \& Abnormal Behavior and Security features simultaneously. Furthermore, 11 tools only provide Question Bank Generation \& Evaluation. Finally, it has been analyzed that most of the proposed tools are not publicly available. This significantly reduces the genuine advantages of proposed tools because further customizations and evaluations are not possible for researchers and practitioners. Further details are given in Section \ref{proposedtool}.
        
        \noindent In addition to the proposed tools, we also identified 25 existing tools (Table \ref{tab10:ExistingTools}) that have been utilized in the selected studies for the implementation of proposals. It has been analyzed that Python and PHP are most frequently used implementation languages in the selected studies. Moreover, MySQL and Firebase are leading storage platforms that have been used in the selected studies. Furthermore, Open CV is most frequently utilized machine learning library in the selected studies. Complete details of utilized tools are available in Section \ref{utilizedtool}.
        
        \item What are the leading datasets proposed / utilized for the online exams solutions? 
        
        \noindent \textbf{Answer:} We overall identified 11 datasets as given in Section \ref{data} (Table \ref{tab11:leadingDatasets}). Six datasets are publicly available whereas the accessibility information of remaining five datasets is unknown. Further details are presented in Section \ref{data}. 
        \item What are the main countries contributed / participated in the online exam research?
        
        \noindent \textbf{Answer:} We overall identified 25 countries along with respective institutes that have participated for online exams research in the selected studies, as given in Table \ref{tab12:ResearchCountry}. It has been analyzed that most of the studies (62\%) belong to Asian countries followed by European countries (17\%). The complete details are available in Section \ref{researchCountry}.
        \item What are the key factors towards the global adoption of online exams and how to promote it in different countries with varying E-learning infrastructure and financial requirements?  
        
        \noindent \textbf{Answer:} On the basis of detailed analysis of selected studies, we identified four most significant factors (i.e. Network Infrastructure, Hardware Requirements, Implementation complexity and training requirements) that significantly influence the global adoption of online exams systems. Particularly, these factors are directly linked with the system’s overall cost, which is a major concerning element for most of the developing countries. To promote the global adoption of online exams, a comparative analysis of these factors with respect to key online exam features is performed (Table \ref{tab13:KeyAdoptionFactors}) in order to investigate the requirements and effects of features on each factor. This facilitates the selection of right online exam system for a particular country on the basis of existing E-learning infrastructure and overall cost. The complete details are given in Section \ref{keyFactors}. 
        \item What are the major challenges in current online exams research and how to improve upon these challenges?  

        \noindent \textbf{Answer:} The investigation of selected studies reveals three major challenges in existing online exams research as follows:
        \begin{enumerate}
            \item \textbf{Inaccessible Tools:} Generally, the proposed tools in literature for a particular problem are publicly available especially in case the study is published in a reputed journal. However, in case of online exams research, the source code and other details of proposed tools are totally inaccessible as highlighted in Section \ref{proposedtool}. This is a major challenge as students, researchers and practitioner are unable to evaluate and / or extend proposed online exam tools. Therefore, the actual benefits of research cannot be achieved. To tackle this challenge, it is essential to propose / develop open source online exams tools where source code and other details are publicly available for further evaluation and extensions.
            \item \textbf{Theoretical Research:} From this SLR, it is analyzed that most of the studies reported in the literature for online exams are theoretical. Particularly, different approaches and frameworks are proposed that are only good for academic purposes and their actual application is highly questionable. Of course, we admit that theoretical aspects are important part of research, however, the domain of online exams demands more practical research that is feasible enough to be deployed in real environment with slight modifications.
            \item \textbf{Economic Requirements:} The area of online exams is highly dependent on the economic situation of a particular country and institute. Therefore, consideration of economic requirements is a significant aspect while proposing some particular online exam solution. In the existing online exams literature, the economic requirements are totally ignored during the proposal of a particular solution. As a result, the proposal becomes infeasible for countries / institutes having low financial conditions even if it is vastly practicable. To tackle this challenge, it is essential to consider economic requirements during the proposal of online exams solution in order to ensure its wide-ranging applications. Similarly, the related aspects of knowledge integration~\cite{91-rashid2014holistic} and quality assurance~\cite{92-rashidSLR} need to be considered in online exams solutions.

        \end{enumerate}
\end{questions}

\subsection{Limitations}
Though, we have carefully followed standard SLR guidelines [10] to carry out this study, few limitations may still be present. For example, we have only considered research studies, which are published in English language. However, there are slight chances that few relevant studies may also exist in languages other than English. Similarly, we have selected six most renowned scientific databases (i.e. IEEE, Elsevier, Springer, ACM, Wiley, and Taylor \& Francis) to carry out this SLR. However, few relevant studies (e.g.~\cite{bawarith2017exam} etc.) may also exist in other scientific repositories as well. In addition to this, we have used several terms (Section \ref{search}) to search relevant studies and subsequently, rejected large number of studies on the basis of paper Title. In this context, there are chances that we initially excluded few relevant studies where title is not reflecting the actual contributions / contents of article.
Despite the aforementioned limitations, the ultimate findings of this SLR are authenticated due to the following reasons: 1) The existence of relevant studies in languages other than English is rare. 2) We have used six most trustworthy databases that usually published peer review high quality studies. Therefore, overall findings of this SLR are reliable and do not change significantly even if few relevant studies are missed from other databases.     

\section{Conclusions and Future Work}
This article presents a Systematic Literature Review (SLR) to identify and investigate 53 studies (published from Jan 2016 to July 2020) pertaining to the online exams domain. This leads to present 5 significant features and 21 proposed tools for online exams. Moreover, 16 important techniques / algorithms and 11 datasets are presented. Furthermore, the participation of 25 countries in online exam research is investigated. Finally, on the basis of SLR results, four key factors for the global adoption of online exams are identified i.e. Network Infrastructure, Hardware Requirements, Implementation Complexity and Training Requirements. Subsequently, the comparative analysis of global adoption factors with significant online exams features is performed. This provides a solid platform for the global adoption of online exams where different countries and institutes can initiate online exams systems on the basis of their existing E-learning infrastructure and overall economic situations.

In future, this study can be extended in multiple directions. For example, one direction is to perform detailed analysis of techniques / algorithms and datasets that are highlighted in this SLR. Another important direction is to develop a complete online exams global adoption system by utilizing the comparison of global factors and key features as given in this article. Particularly, the idea is the development of system where existing situation of global adoption factors is given as an input and the system will predict the feasibility of online exams adoption with optimum features as per given input. We intend to develop such system in our next article.
\section*{Conflict of Interest}
None declared.

\section*{Acknowledgement}
The authors extend their appreciation to the Deanship of Scientific Research at Saudi Electronic University for funding this work under the grant number ELI-CCI20134.



%




\printbibliography

\end{document}